\documentclass[preprint,review,12pt]{elsarticle}

\usepackage{amssymb}
\usepackage{amsmath,bm}
\usepackage{colortbl,booktabs}
\usepackage{tabularx}
\usepackage{threeparttable}
\usepackage{multicol,multirow}
\usepackage{subfigure}
\usepackage [font=footnotesize,labelfont=bf]{caption}
\usepackage{rotating}
\usepackage{tikz}

\usepackage{color}
\definecolor{R}{rgb}{1, 0, 0}
\definecolor{G}{rgb}{0, 0.5, 0}
\definecolor{B}{rgb}{0, 0, 1}

\definecolor{yjzhuCol}{rgb}{0.9, 0, 0.2}
\definecolor{xyhuCol}{rgb}{0.9, 0, 0.9}

\journal{Journal of Computational Physics}

\begin{document}

\begin{frontmatter}

\title{Free-stream preserving linear-upwind and WENO schemes on curvilinear grids}

\author[]{Yujie Zhu}
\author[]{Xiangyu Hu \corref{cor1}}
\cortext[cor1]{Corresponding author}
\address{Department of Mechanical Engineering, Technical University of Munich\\
85748 Graching, Germany}
\ead{xiangyu.hu@tum.de}

\begin{abstract}
	Applying high-order finite-difference schemes, 
	like the extensively used linear-upwind or WENO schemes, 
	to curvilinear grids can be problematic. 
	The geometrically induced error from grid Jacobian and metrics evaluation 
	can pollute the flow field, and degrade the accuracy or cause the simulation failure even 
	when uniform flow imposed, i.e. free-stream preserving problem.
	In order to address this issue, a method for general linear-upwind and WENO schemes preserving free-stream 
	on stationary curvilinear grids is proposed. 
	Following Lax-Friedrichs splitting, this method rewrites the numerical flux into a central term, 
	which achieves free-stream preserving by using symmetrical conservative metric method,  
	and a numerical dissipative term with a local difference form of 
	conservative variables for neighboring grid-point pairs. 
	In order to achieve free-stream preservation for the latter term, 
	the local difference are modified to share the same Jacobian and metric terms evaluated by high order schemes. 
	In addition, a simple hybrid scheme switching between linear-upwind and WENO schemes
	is proposed of improving computational efficiency and reducing numerical dissipation. 
	A number of testing cases including free-stream, isentropic vortex convection, double Mach reflection 
	and flow past a cylinder are computed to verify the effectiveness of this method.
\end{abstract}


\end{frontmatter}


\section{Introduction}
\label{sec1}
In computational fluid dynamics (CFD), it is well known that finite-difference schemes 
are more computational efficient and easier to achieve high-order accuracy compared 
with finite-volume schemes~\cite{shu2003high}. 
Therefore, many linear, nonlinear
and hybrid high-order finite-different schemes have been developed.
However, despite the above-mentioned advantages, these high-order schemes 
are problematic when they are applied to curvilinear grids due to 
the lack of geometric conservation law (GCL)~\cite{thomas1979geometric,vinokur1989analysis}.
The grid Jacobian and metrics calculated in curvilinear coordinates 
can introduce large errors, degrade accuracy or cause numerical instability
even when the flow is uniform, i.e. free-stream preserving problem.
Since body-fitted curvilinear grids are widely used for computing flow problems 
involving practical geometries, GCL is of great importance.
GCL comprises two components, 
i.e. the volume conservation law (VCL) relevant to moving grid 
and the surface conservation law (SCL) to stationary curvilinear grid. 
While violating VCL causes non-physical extra source or sinks, 
violating SCL leads to a misrepresentation of the convective velocities,
which can be explained as inconsistence of vectorized computational cell surfaces 
in a finite-volume point of view~\cite{zhang1993discrete, vinokur1989analysis, deng2013further}. 
Here, we only consider SCL of stationary grid on which VCL is automatically satisfied.

For low-order schemes,
SCL can be achieved by simple averaging method~\cite{pulliam1980implicit},
conservative form of metrics~\cite{thomas1979geometric} or 
finite-volume-like technique~\cite{vinokur1989analysis}.
It has also been shown that SCL can be satisfied for high-order central 
schemes which are characterized by non-dissipation property. 
In a study of central compact scheme ~\cite{lele1992compact} on curvilinear grid, 
Visbal and Gaitonde~\cite{Visbal2002} found that the SCL error can be largely decreased 
by a two-step procedure: 
(a) utilizing the conservative form of metrics~\cite{thomas1979geometric},
(b) discretizing the metric terms with the same compact scheme which is used 
for calculating the convective-flux derivatives.
In a further analysis, Deng et al.~\cite{deng2011geometric} 
identified the outer- and inner-level
differential operators for the metrics, and obtained 
a sufficient SCL condition for general high-order central schemes
~\cite{deng2013further, abe2013short}.
However, as also pointed in Refs.~\cite{Nonomura2010, deng2013further},
this condition is difficult to be satisfied for dissipative, i.e. upwind schemes
due to the inconsistent outer-level differential operators used for flux splitting.

Since numerical dissipation is essential for stabilizing the solution, 
especially for compressible flow with shock, such difficulty 
is usually circumvented by the combination of finite-difference 
and finite-volume schemes. 
One typical formulation is first obtaining, usually nonlinear, 
dissipative convective-fluxes by finite-volume approach and 
then computing their derivatives by applying non-dissipative central schemes
~\cite{Nonomura2010, deng2013further, jiang2013alternative}.
Note that this formulation is in agreement with 
that of the original weighted compact nonlinear scheme (WCNS)
developed by Deng and Zhang~\cite{deng2000developing},
in which the convective-fluxes is computed with the finite-volume version 
of a weighted essentially non-oscillatory (WENO) scheme~\cite{jiang1996efficient}.
Another formulation is first split the upwind scheme into a non-dissipative central part 
and a dissipative part, and then implementing them, respectively, 
with high-order finite-difference and finite-volume-like schemes 
by freezing Jacobian and metric terms for the entire stencil
~\cite{nonomura2015new} or by replacing the transformed conservative variables with the original ones~\cite{vinokur2002extension, li2017further}.
Recently, a finite-difference based free-stream preserving technique  
was proposed by Zhu et al.~\cite{zhu2017numerical} for WENO scheme.
In this technique, the consistency of outer-level difference operators
is imposed by introducing offsetting terms with the same WENO nonlinear weights 
for computing the corresponding inviscid fluxes.
While this technique is generally effective, 
it may lead to large errors due to the resulting non-conservative formulation.

In this work, 
we propose a simple technique to impose SCL for linear-upwind 
and WENO schemes and their hybridization to achieve free-stream preserving property. 
To our knowledge, this is the first report on hybrid WENO scheme on curvilinear grids 
with free-stream preserving property.
The method follows a Lax-Friedrichs splitting 
to rewrite the numerical flux of a upwind scheme into a central term of the flux functions 
and a dissipative term.
Since the latter term is transformed into a formulation of local differences 
for neighboring grid-point pairs, the Jacobian and metric terms can be 
evaluated with high-order schemes and applied for each of these pairs.
The reminder of the paper is organized as follows. 
Section 2 describes the classic formulation of central schemes satisfying SCL, 
linear-upwind and WENO scheme on curvilinear grids based 
on Lax-Friedrichs splitting and their difficult on free-stream preserving.
In Section 3, the local-difference formulations of the dissipative term 
for linear-upwind and WENO schemes are introduced 
with a further application of hybrid WENO scheme following Hu et al. ~\cite{hu2015efficient}.
Validation tests and further numerical examples are presented in Section 4, 
and brief concluding remarks are given in the last Section 5.

\section{Preliminaries}
\label{sec1}
In Cartesian coordinates $ \left( {t, x,y,z} \right)$, 
the three-dimensional Euler equation is given as follows:

\begin{equation}
\frac{\partial \bm{U}}{{\partial t}} +
\frac{\partial \bm{F}}{{\partial x}}
+ \frac{\partial \bm{G}}{{\partial y}}
+ \frac{\partial \bm{H}}{{\partial z}} =0,
\label{goveq}
\end{equation}
where 
\begin{equation}
\begin{split}
& \bm{U} = {\left(  {\begin{array}{*{20}{c}}
		\rho  & {\rho u} & {\rho v} & {\rho w} & E  \\
		\end{array}} \right) ^T}, \\
& \bm{F} = {\left(  {\begin{array}{*{20}{c}}
		{\rho u}  & {\rho u^2+p} & {\rho uv} & {\rho uw} & {u\left( E+p\right) }  \\
		\end{array}} \right) ^T} ,  \\
& \bm{G} = {\left(  {\begin{array}{*{20}{c}}
		{\rho v}  & {\rho uv} & {\rho v^2+p} & {\rho vw} & {v\left( E+p\right) }  \\
		\end{array}} \right) ^T},   \\
& \bm{H} = {\left(  {\begin{array}{*{20}{c}}
		{\rho w}  & {\rho uw} & {\rho vw} &{\rho w^2+p}  & {w\left( E+p\right) }  \\
		\end{array}} \right) ^T} ,
\end{split} 
\label{eq2}
\end{equation}
are conservative variables and convective fluxes, respectively. 
Here, $ \rho $  is density; $ p $ is pressure; 
$ {u, v, w} $ denote the velocity components in $ {x-} $,$ {y-} $and $ {z-} $directions, respectively; 
and $ {E} $ is total energy per unit volume. 
In this study, the ideal gas is used and $ {E} $ can be expressed as
\begin{equation}
E=\frac{p}{{\gamma -1}}+
\frac{1}{{2}} \rho \left( u^2+v^2+w^2\right) .
\label{energy}
\end{equation}

\subsection{Governing equations in curvilinear coordinates}
\label{subsec1}
When a curvilinear grid is used for numerical discretization, 
the governing equation Eq.~\eqref{goveq} is first transformed into 
curvilinear coordinates $ {\left(\tau, \xi, \eta, \zeta\right) } $ 
with the following relationships
\begin{equation}
\tau = t,  \quad
\xi = \xi\left( x, y ,z\right),  \quad
\eta= \eta \left( x, y, z\right) ,   \quad
\zeta= \zeta\left( x, y, z\right) .
\label{eq4}
\end{equation}
The transformed equation can be written as
\begin{equation}
\frac{\partial \bm{\tilde{U}}}{{\partial \tau}} +
\frac{\partial \bm{\tilde{F}}}{{\partial \xi}}
+ \frac{\partial \bm{\tilde{G}}}{{\partial \eta}}
+ \frac{\partial \bm{\tilde{H}}}{{\partial \zeta}} =0,
\label{goveq1}
\end{equation}
where
\begin{equation}
\begin{split}
& \bm{\tilde{U}}=\frac{\bm{U}}{J}, \\ 
& \bm{\tilde{F}}=\frac{{{\xi }_{x}}}{J}\bm{F}+\frac{{{\xi }_{y}}}{J}\bm{G}+\frac{{{\xi }_{z}}}{J}\bm{H}, \\ 
& \bm{\tilde{G}}=\frac{{{\eta }_{x}}}{J}\bm{F}+\frac{{{\eta }_{y}}}{J}\bm{G}+\frac{{{\eta }_{z}}}{J}\bm{H}, \\ 
& \bm{\tilde{H}}=\frac{{{\zeta }_{y}}}{J}\bm{F}+\frac{{{\zeta }_{y}}}{J}\bm{G}+\frac{{{\zeta }_{z}}}{J}\bm{H}.
\end{split} 
\label{eq6}
\end{equation}
Here, the transformation Jacobian $J$ and metrics are
\begin{equation}
\begin{split}
& \frac{1}{J}=x_\xi y_\eta z_\zeta-x_\eta y_\xi z_\zeta+x_\zeta y_\xi z_\eta -x_\xi y_\zeta z_\eta+x_\eta y_\zeta z_\xi-x_\zeta y_\eta z_\xi, \\ 
& \frac{{{\xi }_{x}}}{J}=y_\eta z_\zeta-y_\zeta z_\eta,
\quad \frac{{{\xi }_{y}}}{J}=x_\zeta z_\eta-x_\eta z_\zeta,
\quad \frac{{{\xi }_{z}}}{J}=x_\eta y_\zeta-x_\zeta y_\eta,\\
&\frac{{{\eta }_{x}}}{J}=y_\zeta z_\xi-y_\xi z_\zeta,
\quad \frac{{{\eta }_{y}}}{J}=x_\xi z_\zeta-x_\zeta z_\xi,
\quad \frac{{{\eta }_{z}}}{J}=x_\zeta y_\xi-x_\xi y_\zeta,\\
& \frac{{{\zeta }_{y}}}{J}=y_\xi z_\eta-y_\eta z_\xi,
\quad \frac{{{\zeta }_{y}}}{J}=x_\eta z_\xi-x_\xi z_\eta,
\quad \frac{{{\zeta }_{z}}}{J}=x_\xi y_\eta-x_\eta y_\xi,
\end{split} 
\label{metrics0}
\end{equation}
and the equation Jacobian matrix, say $\bm{A}=\partial \bm{\tilde{F}}/\partial \bm{\tilde{U}}$, is
	
\begin{equation}
\bm{A}=
\left[ \begin{array}{ccccc}
0    &\xi_x   &\xi_y    &\xi_z   &0    \\
\xi_x \phi-u \theta &\theta-\left( \gamma-2\right)u\xi_x &u\xi_y-\left( \gamma-1\right)v\xi_x &u\xi_z-\left( \gamma-1\right)w\xi_x &\left( \gamma-1\right)\xi_x\\
\xi_y \phi-v \theta &v\xi_x-\left( \gamma-1\right)u\xi_y &\theta-\left( \gamma-2\right)v\xi_y &v\xi_z-\left( \gamma-1\right)w\xi_y &\left( \gamma-1\right)\xi_y\\
\xi_z \phi-w \theta &w\xi_x-\left( \gamma-1\right)u\xi_z &w\xi_y-\left( \gamma-1\right)v\xi_z &\theta-\left( \gamma-2\right)w\xi_z &\left( \gamma-1\right)\xi_z\\
\left( \phi-h\right)\theta&H\xi_x-\left(\gamma-1 \right)u\theta& H\xi_y-\left(\gamma-1 \right)v\theta&  H\xi_z-\left(\gamma-1 \right)w\theta& \gamma\theta
\end{array} \right], 
\label{matirx-A}
\end{equation}	
where
\begin{equation}
\begin{split}
& \theta=\frac{\gamma-1}{2}\left(u^2+v^2+w^2 \right),  \\ 
& \theta=\xi_x u+\xi_y v+\xi_z w,\\
& H=\frac{\gamma p}{\left( \gamma-1\right)\rho }+\frac{1}{2}\left( u^2+v^2+w^2\right). \\
\end{split} 
\label{matrix-A-1}
\end{equation}	
The eigenvalues of $\bm{A}$ are
	\begin{equation}
	\begin{split}
	& \lambda_1=u \xi_x+v \xi_y+w \xi_z-a\sqrt{\xi_x^2+\xi_y^2+\xi_z^2},  \\ 
	& \lambda_2=\lambda_3=\lambda_4=u \xi_x+v \xi_y+w \xi_z,\\
	& \lambda_5=u \xi_x+v \xi_y+w \xi_z+a\sqrt{\xi_x^2+\xi_y^2+\xi_z^2}, \\
	\end{split} 
	\label{matrix-A-2}
	\end{equation}
where $a=\sqrt{\gamma p/\rho}$ is the speed of sound.

\subsection{SCL and free-stream preserving problem}
\label{subsec1}
When a uniform flow is imposed, Eq.~\eqref{goveq1} can be simplified as
\begin{equation}
\bm{\tilde{U}}_\tau =-\left( I_x \bm{F}+I_y \bm{G} +I_z \bm{H}\right) = 0,
\label{goveq2}
\end{equation}
where
\begin{equation}
\begin{split}
& I_x =\left( \frac{{{\xi }_{x}}}{J}\right)_\xi+\left( \frac{{{\eta }_{x}}}{J}\right)_\eta 
+\left( \frac{{{\zeta }_{x}}}{J}\right)_\zeta = 0, \\
& I_y=\left( \frac{{{\xi }_{y}}}{J}\right)_\xi+\left( \frac{{{\eta }_{y}}}{J}\right)_\eta
+\left( \frac{{{\zeta }_{y}}}{J}\right)_\zeta = 0,\\
& I_z=\left( \frac{{{\xi }_{z}}}{J}\right)_\xi+\left( \frac{{{\eta }_{z}}}{J}\right)_\eta
+\left( \frac{{{\zeta }_{z}}}{J}\right) _\zeta = 0.
\end{split} 
\label{eq9}
\end{equation}
Note that, Eq. (\ref{eq9}) may not be strictly satisfied when $I_x$, $I_y$ and $I_z$ 
are represented by numerical discretization.   
In this case, artificial numerical disturbances may be introduced into the uniform flow 
and lead to the free-stream preserving problem. 
As Eq. (\ref{eq9}) can be explained as the consistence of vectorized computational cell surfaces 
in finite-volume method~\cite{vinokur1989analysis},
Zhang et al.~\cite{zhang1993discrete} proposed 
the surface conservation law (SCL),
by which Eq. (\ref{eq9}) is still satisfied by the numerical 
approximation of Jacobian, metrics and their derivative operators.
It is obvious that a numerical scheme satisfying SCL has the free-stream preserving property.  

Following Deng et al.~\cite{deng2013further}, 
the metric terms in Eq.~\eqref{metrics0} is rewritten 
as a symmetrical conservative form:
\begin{equation}
	\begin{split}
		& \frac{{{\xi }_{x}}}{J}=\frac{1}{2}\left[ \left( y_\eta z\right)_\zeta -\left( y_\zeta z\right) _\eta +\left( y z_\zeta\right)_\eta -\left( y z_\eta\right) _\zeta\right] ,\\
		&\frac{{{\xi }_{y}}}{J}=\frac{1}{2}\left[ \left( x z_\eta\right) _\zeta-\left( x z_\zeta\right) _\eta +\left( x_\zeta z\right) _\eta-\left( x_\eta z\right) _\zeta\right] ,\\
		& \frac{{{\xi }_{z}}}{J}=\frac{1}{2}\left[\left( x_\eta y\right) _\zeta-\left( x_\zeta y\right) _\eta +\left( x y_\zeta\right) _\eta-\left( x y_\eta\right) _\zeta\right],\\
		& \frac{{{\eta }_{x}}}{J}  =\frac{1}{2}\left[\left( y_\zeta z\right) _\xi-\left( y_\xi z\right) _\zeta +\left( y z_\xi\right) _\zeta-\left( y z_\zeta\right) _\xi\right] , \\
		& \frac{{{\eta }_{y}}}{J} =\frac{1}{2}\left[ \left( x z_\zeta\right) _\xi-\left( x z_\xi\right) _\zeta+ \left( x_\xi z\right) _\zeta-\left( x_\zeta z\right) _\xi\right] , \\
		& \frac{{{\eta }_{z}}}{J} =\frac{1}{2}\left[ \left( x_\zeta y\right) _\xi-\left( x_\xi y\right) _\zeta +\left( x y_\xi\right) _\zeta-\left( x y_\zeta\right) _\xi\right] , \\
		& \frac{{{\zeta }_{x}}}{J}=\frac{1}{2}\left[ \left( y_\xi z\right) _\eta-\left( y_\eta z\right) _\xi +\left( y z_\eta\right) _\xi-\left( y z_\xi\right) _\eta\right] ,\\
		& \frac{{{\zeta }_{y}}}{J}=\frac{1}{2}\left[ \left( x z_\xi\right) _\eta-\left( x z_\eta\right) _\xi +\left( x_\eta z\right) _\xi-\left( x_\xi z\right) _\eta\right] ,\\
		& \frac{{{\zeta }_{z}}}{J}=\frac{1}{2}\left[ \left( x_\xi y\right) _\eta-\left( x_\eta y\right) _\xi +\left( x y_\eta\right) _\xi-\left( x y_\xi\right) _\eta\right]
	\end{split} 
	\label{metrics2}
\end{equation}
and 
\begin{equation}
	\begin{split}
		\frac{1}{J}=\frac{1}{3}\left[ \left( x \frac{\xi_{x}}{J}+y \frac{\xi_{y}}{J}+z \frac{\xi_{z}}{J} \right)_{\xi}+ \left( x \frac{\eta_{x}}{J}+y \frac{\eta_{y}}{J}+z \frac{\eta_{z}}{J} \right)_{\eta}+\left(x \frac{\zeta_{x}}{J}+y \frac{\zeta_{y}}{J}+z \frac{\zeta_{z}}{J} \right)_{\zeta}\right]. 
	\end{split}
	\label{eq13}
\end{equation}
As shown in Refs.~\cite{Visbal2002,Nonomura2010,deng2011geometric}, 
when the derivative operators within the above conservative form 
are kept the same with that of fluxes, 
Eq. (\ref{eq9}) is satisfied and uniform flow can be preserved. 
Taking $ I_x $ as an example,
\begin{equation}
\begin{split}
{I_x}=\frac{1}{2}\left[ {\delta_1}^{\xi}{\delta_2}^{\zeta}\left( z {\delta_3}^{\eta}y\right) -{\delta_1}^{\xi}{\delta_2}^{\eta}\left( z {\delta_3}^{\zeta}y\right)+ 
{\delta_1}^{\xi}{\delta_2}^{\eta}\left( y {\delta_3}^{\zeta}z\right) -{\delta_1}^{\xi}{\delta_2}^{\zeta}\left( y {\delta_3}^{\eta}z\right) \right. \\ +{\delta_1}^{\eta}{\delta_2}^{\xi}\left( z {\delta_3}^{\zeta}y\right) -{\delta_1}^{\eta}{\delta_2}^{\zeta}\left( z {\delta_3}^{\xi}y\right) 
+{\delta_1}^{\eta}{\delta_2}^{\zeta}\left( y {\delta_3}^{\xi}z\right) -{\delta_1}^{\eta}{\delta_2}^{\xi}\left( y {\delta_3}^{\zeta}z\right)\\ \left.+{\delta_1}^{\zeta}{\delta_2}^{\eta}\left( z {\delta_3}^{\xi}y\right) -{\delta_1}^{\zeta}{\delta_2}^{\xi}\left( z {\delta_3}^{\eta}y\right)
+{\delta_1}^{\zeta}{\delta_2}^{\xi}\left( y {\delta_3}^{\eta}z\right) -{\delta_1}^{\zeta}{\delta_2}^{\eta}\left( y {\delta_3}^{\xi}z\right)\right].
\end{split}
\label{eq11}
\end{equation}   
Here, $ \delta_1 $, $ \delta_2 $ are outer derivative operators 
and $ \delta_3 $ is inner derivative operator for calculating 
the corresponding level of the metric terms. 
The superscript $ \xi $, $ \eta $ and $ \zeta $ denote the operators 
in $ \xi- $, $ \eta- $ and $ \zeta- $ directions, respectively. 
It is straightforward to see that $ I_x $ equals to zero when ${\delta_2}^{\xi}={\delta_1}^{\xi}$, 
${\delta_2}^{\eta}={\delta_1}^{\eta}$ and ${\delta_2}^{\zeta}={\delta_1}^{\zeta}  $. 
This technique is called symmetrical conservative metric method (SCMM) in Ref.~\cite{deng2013further} 
and is effective for central schemes. 
However, as will be shown in the following subsection,
this method is difficult to be applied for upwind schemes. 
	
\subsection{Linear-upwind scheme}
\label{subsec1}
Without loss of generality, we explain the explicit 5th-order linear upwind scheme 
with local Lax-Friedrichs splitting.

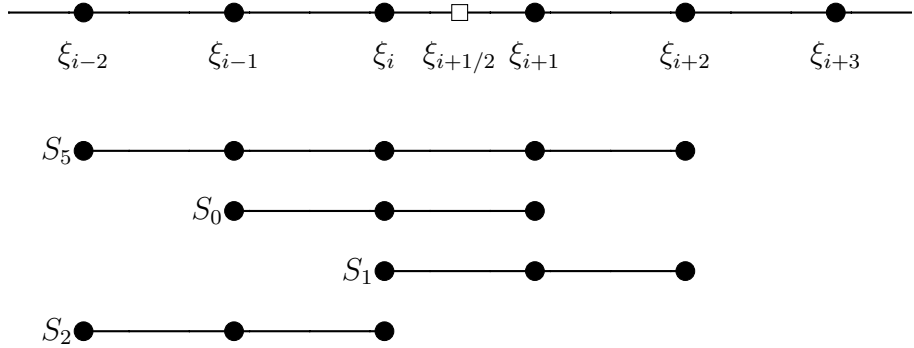
\begin{figure}[tbh]
	\begin{center}
		\begin{tikzpicture}[
		dot/.style 2 args={circle,draw=#1,fill=#2,inner sep=2.5pt},
		square/.style 2 args={draw=#1,fill=#2,inner sep=3pt},
		mystar/.style 2 args={star,draw=#1,fill=#2,inner sep=1.5pt},
		mydiamond/.style 2 args={diamond,draw=#1,fill=#2,inner sep=1.5pt},
		scale=0.8
		]
		
		\draw[black,thick]  (0,-1) grid (15,-1);
		\draw[black,thick,yshift=-0.3cm]  (1.25,-3) grid (11.25,-3);
		\draw[black,thick,yshift=-0.3cm]  (3.75,-4) grid (8.75,-4);
		\draw[black,thick,yshift=-0.3cm]  (6.25,-5) grid (11.25,-5);
		\draw[black,thick,yshift=-0.3cm]  (1.25,-6) grid (6.25,-6);
		\foreach \Fila in {1.25,3.75,6.25,8.75,11.25,13.75}{\node[dot={black}{black}] at (\Fila,-1) {};}  
		\foreach \Fila in {7.5}{\node[square={black}{white}] at (\Fila,-1) {};}  
		\foreach \Fila in {1.25,3.75,6.25,8.75,11.25}{\node[dot={black}{black}] at (\Fila,-3.3) {};} 
		\foreach \Fila in {3.75,6.25,8.75}{\node[dot={black}{black}] at (\Fila,-4.3) {};}  
		\foreach \Fila in {6.25,8.75,11.25}{\node[dot={black}{black}] at (\Fila,-5.3) {};}   
		\foreach \Fila in {1.25,3.75,6.25}{\node[dot={black}{black}] at (\Fila,-6.3) {};}  
		\node[below] at (1.25,-1.3) {$\xi_{i-2}$};
		\node[below] at (3.75,-1.3) {$\xi_{i-1}$};
		\node[below] at (6.25,-1.3) {$\xi_{i}$};
		\node[below] at (8.75,-1.3) {$\xi_{i+1}$};
		\node[below] at (11.25,-1.3) {$\xi_{i+2}$};
		\node[below] at (13.75,-1.3) {$\xi_{i+3}$};
		\node[below] at (7.5,-1.3) {$\xi_{i+1/2}$};
		\node[left]  at (1.25,-3.3) {$S_{5}$};
		\node[left]  at (3.75,-4.3) {$S_{0}$};
		\node[left]  at (6.25,-5.3) {$S_{1}$};
		\node[left]  at (1.25,-6.3) {$S_{2}$};
		\end{tikzpicture}
	\end{center}		
	\caption{Full stencil and candidate stencils for constructing 
		the positive characteristic fluxes at the half point $\xi_{i+1/2}$. 
		Here, the stencil $S_5$ is used for the 5th-order linear-upwind scheme 
		and the stencils $S_0$, $S_1$, $S_2$ for WENO scheme. }
	\label{stencil}	
\end{figure}

The semi-discrete approximation of the governing equation Eq.~\eqref{goveq1} 
at a grid point indexed as $(i, j, k)$ is as follows:
\begin{equation}
\begin{split}
\left( \frac{\partial \bm{\tilde{U}}}{{\partial \tau}}\right)_{i,j,k} =-\left( 
\delta_1^{\xi}{ \bm{\tilde{F}}}_{i,j,k}
+ \delta_1^{\eta}{ \bm{\tilde{G}}}_{i,j,k}
+ \delta_1^{\zeta}{ \bm{\tilde{H}}}_{i,j,k}\right),
\end{split}
\label{eq14}
\end{equation}
where $ \delta_1^{\xi} $, $ \delta_1^{\eta} $ and $ \delta_1^{\zeta} $ 
are flux derivative operators in $ {\xi}- $, $ {\eta}- $ and $ {\zeta}- $ directions, respectively. 
A conservative formulation of the operators, say $ \delta_1^{\xi}{ \bm{\tilde{F}}}_{i,j,k} $, is 
\begin{equation}
\begin{split}
\delta_1^{\xi}{ \bm{\tilde{F}}}_{i,j,k}=\frac{1}{\delta \xi}\left( { \bm{\tilde{F}}}_{i+\frac{1}{2},j,k}-{ \bm{\tilde{F}}}_{i-\frac{1}{2},j,k}\right) ,
\end{split}
\label{eq15}
\end{equation}
where ${\bm{\tilde{F}}}_{i\pm\frac{1}{2},j,k}$ are the numerical fluxes at half points. 
Here, $\delta \xi$ is the equidistant space step and is selected as follows in this work,
 \begin{equation}
 \begin{split}
 \delta{\xi}=\frac{L_{\xi}}{N_{\xi}} ,
 \end{split}
 \label{eqdelta-xi}
 \end{equation}
 where $L_{\xi}$ and $N_{\xi}$ are the characteristic length scale 
and the grid number in $\xi-$ direction, respectively.
The detailed calculating procedure, say for ${\bm{\tilde{F}}}_{i+\frac{1}{2}}$, 
includes the following steps. Here, since a one-dimensional stencil is used, 
as shown in Fig. \ref{stencil}, 
the subscript $ j $ and $ k $ are omitted in this section for simplicity.
First, transform the fluxes and conservative variables at all grid-points within the stencil 
into characteristic space and carry out a local Lax-Friedrichs splitting:
\begin{equation}
\begin{split}
f_m^{s,\pm}=\frac{1}{2} \bm{L}_{i+\frac{1}{2}}^{s} \cdot \left(\bm{\tilde{F}}_m \pm \lambda^{s} \bm{\tilde{U}}_m\right)
\end{split} \quad m=i-2,i+3,
\label{eq16}
\end{equation}
where $ f_m^{s,\pm} $ denotes the $ s- $th positive and negative characteristic fluxes, 
$ \bm{L}_{i+\frac{1}{2}}^{s} $ is the $ s- $th left eigenvector vector of   
the linearized Roe-average Jacobian matrix 
$ \bm{A}_{i+1/2}= \left(\partial \bm{\tilde{F}}/{\partial \bm{\tilde {U}}}\right)_{i+1/2}$~\cite{roe1981approximate}  
and $\lambda^{s}=\max\left( \left| \lambda_m^{s}\right| \right)$ denotes the largest $ s- $th 
eigenvalue of the Jacobian $ \bm{A}$ across the stencil.
Then, construct the characteristic fluxes at the half-point as follows:
\begin{equation}
\begin{split}
&f_{i+\frac{1}{2}}^{s,+}=\frac{1}{60} \left( 2f_{i-2}^{s,+}-13f_{i-1}^{s,+}+47f_{i}^{s,+}+ 27f_{i+1}^{s,+}-3f_{i+2}^{s,+}\right), \\
&f_{i+\frac{1}{2}}^{s,-}=\frac{1}{60} \left( -3f_{i-1}^{s,-}+27f_{i}^{s,-}+47f_{i+1}^{s,-} -13f_{i+2}^{s,-}+2f_{i+3}^{s,-}\right).
\end{split} 
\label{eq18}
\end{equation}
Finally, transform the characteristic fluxes back into physical space by
\begin{equation}
{ \bm{\tilde{F}}}_{i+\frac{1}{2}}=
\sum_{s}\bm{R}_{i+\frac{1}{2}}^{s} \left({f}_{i+\frac{1}{2}}^{s,+} + {f}_{i+\frac{1}{2}}^{s,-}\right),
\label{eq19}
\end{equation}
where $\bm{R}_{i+\frac{1}{2}}^{s}$ is the s-th right eigenvector vector of 
$ \bm{A}_{i+1/2}$. Substituting Eqs.~\eqref{eq16} and~\eqref{eq18} into Eq.~\eqref{eq19}, 
the numerical flux can be expressed as
\begin{equation}
\begin{split}
&{\bm{\tilde{F}}}_{i+\frac{1}{2}}
=\frac{1}{60}\left( \bm{\tilde{F}}_{i-2}- 8\bm{\tilde{F}}_{i-1} +37\bm{\tilde{F}}_{i} +37\bm{\tilde{F}}_{i+1} -8\bm{\tilde{F}}_{i+2}+\bm{\tilde{F}}_{i+3} \right) \\
&\hspace{8mm}+\frac{1}{60}\sum_{s}\bm{R}_{i+\frac{1}{2}}^s  {\lambda}^s  \bm{L}_{i+\frac{1}{2}}^s \cdot \left( \bm{\tilde{U}}_{i-2}- 5\bm{\tilde{U}}_{i-1} +10\bm{\tilde{U}}_{i} -10\bm{\tilde{U}}_{i+1} +5\bm{\tilde{U}}_{i+2}-\bm{\tilde{U}}_{i+3} \right).
\end{split} 
\label{eq20}
\end{equation}
Note that, the numerical flux contains both the fluxes 
and the conservative variables $\bm{\tilde{U}}$ term. 
Due to the extra term of conservative variables $\bm{\tilde{U}}$, 
the derivatives of operator $ \delta_1 $ in Eq.~\eqref{eq11} is 
an inconsistent operator and  making operator $ \delta_2 $ 
equal to $ \delta_1 $ is difficult or impossible. 
Therefore, the symmetrical conservative metric method is not valid for linear-upwind scheme 
and a uniform flow is not preserved.

\subsection{WENO scheme}
\label{subsec1}
In the typical 5th-order WENO scheme ~\cite{jiang1996efficient} 
with local Lax-Friedrichs flux splitting as given in Eq. (\ref{eq16}),
the positive WENO characteristic flux can be expressed as 
\begin{equation}
f_{i+\frac{1}{2}}^{s,+}=\sum\limits_{k=0}^{2}\omega_k^{+}q_k^{+}, \\
\label{eq24}
\end{equation}
where
\begin{equation}
\begin{split}
&q_0^{+}=\frac{1}{3}f_{i-2}^{s,+}-\frac{7}{6}f_{i-1}^{s,+}+\frac{11}{6}f_{i}^{s,+}, \\
&q_1^{+}=-\frac{1}{6}f_{i-1}^{s,+}+\frac{5}{6}f_{i}^{s,+}+\frac{1}{3}f_{i+1}^{s,+}, \\
&q_2^{+}=\frac{1}{3}f_{i}^{s,+}+\frac{5}{6}f_{i+1}^{s,+}-\frac{1}{6}f_{i+2}^{s,+}, 
\end{split} 
\label{eq25}
\end{equation} 
are 3rd-order approximations using the stencils as shown in Fig. \ref{stencil}. 
$\omega_k^{+}$ in Eq. (\ref{eq24}) are the corresponding nonlinear weights determined by
\begin{equation}
\omega_k^{+}=\left. \frac{C_k^{+}}{\left( \varepsilon+\beta_k^{+}\right)^{2}} 
\middle/ \sum\limits_{r=0}^{2}\frac{C_r^{+}}{\left( \varepsilon+\beta_r^{+}\right)^{2}} \right., \\
\label{eq26}
\end{equation}
where 
\begin{equation}
\begin{split}
&C_0^{+}=\frac{1}{10},\quad C_1^{+}=\frac{3}{5},\quad C_2^{+}=\frac{3}{10}, \\
& \beta_0^{+}=\frac{1}{4}\left(f_{i-2}^{s,+}-4f_{i-1}^{s,+}+3f_{i}^{s,+} \right)^2+\frac{13}{12}\left(f_{i-2}^{s,+}-2f_{i-1}^{s,+}+f_{i}^{s,+} \right)^2,\\
&\beta_1^{+}=\frac{1}{4}\left(-f_{i-1}^{s,+}+f_{i+1}^{s,+} \right)^2+\frac{13}{12}\left(f_{i-1}^{s,+}-2f_{i}^{s,+}+f_{i+1}^{s,+} \right)^2,\\
&\beta_2^{+}=\frac{1}{4}\left(-3f_{i}^{s,+}+4f_{i+1}^{s,+}-f_{i+2}^{s,+} \right)^2+\frac{13}{12}\left(f_{i}^{s,+}-2f_{i+1}^{s,+}+f_{i+2}^{s,+} \right)^2,
\end{split} 
\label{eq27}
\end{equation}
are the optimal weights for a background linear-upwind scheme 
and smoothness indicators of the corresponding stencil, respectively. 
$\varepsilon = 10^{-6}$ is a small positive parameter to prevent division by zero.
Note that, the negative fluxes $f_{i+\frac{1}{2}}^{s,-}$ can be obtained in a similar way 
by flipping the stencils respect to $\xi_{i+1/2}$. 
Then, the numerical fluxes at the half point in physical space are constructed by
\begin{equation}
\begin{split}
&{ \bm{\tilde{F}}}_{i+\frac{1}{2}}=\sum_{s}\bm{R}_{i+\frac{1}{2}}^s \left( {f}_{i+\frac{1}{2}}^{s,+}+{f}_{i+\frac{1}{2}}^{s,-}\right).
\end{split} 
\label{eqweno}
\end{equation}

In Refs.~\cite{jiang1999high,nonomura2015new}, 
the WENO fluxes are divided into a central part and a dissipation part as follows:
\begin{equation}
\begin{split}
&{ \bm{\tilde{F}}}_{i+\frac{1}{2}}={ \bm{\tilde{F}}}_{i+\frac{1}{2}}^{+}+{ \bm{\tilde{F}}}_{i+\frac{1}{2}}^{-} \\
&=\sum_{s}\bm{R}_{i+\frac{1}{2}}^s {f}_{i+\frac{1}{2}}^{s,+}+ \sum_{s}\bm{R}_{i+\frac{1}{2}}^s  {f}_{i+\frac{1}{2}}^{s,-}\\
& =\frac{1}{60}\left( \bm{\tilde{F}}_{i-2}- 8\bm{\tilde{F}}_{i-1} +37\bm{\tilde{F}}_{i} +37\bm{\tilde{F}}_{i+1} -8\bm{\tilde{F}}_{i+2}+\bm{\tilde{F}}_{i+3} \right) \\
& -\frac{1}{60}\sum_{s}\bm{R}_{i+\frac{1}{2}}^s  \left\lbrace \left( 20\omega_1^{+}-1\right) {\widehat{f}}_{i,1}^{s,+}-\left( 10\left(\omega_1^{+}+\omega_2^{+} \right) -5\right){\widehat{f}}_{i,2}^{s,+}+{\widehat{f}}_{i,3}^{s,+}  \right\rbrace \\
& +\frac{1}{60}\sum_{s}\bm{R}_{i+\frac{1}{2}}^s  \left\lbrace \left( 20\omega_1^{-}-1\right) {\widehat{f}}_{i,1}^{s,-}-\left( 10\left(\omega_1^{-}+\omega_2^{-} \right) -5\right){\widehat{f}}_{i,2}^{s,-}+{\widehat{f}}_{i,3}^{s,-}  \right\rbrace, \\
\end{split} 
\label{eq28}
\end{equation}
where 
\begin{equation}
\begin{split}
&\widehat{f}_{i,r+1}^{s,+}=f_{i+r+1}^{s,+}-3f_{r}^{s,+}+3f_{i+r-1}^{s,+}-f_{i+r-2}^{s,+}, \quad r = 0, 1, 2,\\
&\widehat{f}_{i,r+1}^{s,-}=f_{i-r+3}^{s,-}-3f_{i-r+2}^{s,-}+3f_{i-r+1}^{s,-}-f_{i-r}^{s,-}, \quad r = 0, 1, 2.\\
\end{split} 
\label{eq29}
\end{equation}
Similar to the linear-upwind scheme, the operator $ \delta_1 $ 
in WENO scheme is also an inconsistent operator because of the fluxes splitting. 
Since the nonlinear weighted numerical flux depends on the smoothness of each stencil,
it leads to extra inconsistency between $ \delta_2 $ and $ \delta_1 $.      
Therefore, it is difficult to use the symmetrical conservative metric method 
to achieve free-stream preserving property for WENO scheme. 
Another issue is that the approximation of Jacobian and metrics 
also introduces disturbances into the smoothness indicators of Eq. \eqref{eq27}, 
so that the optimal background linear scheme is not recovered for a uniform flow.

\section{Free-stream preserving upwind schemes}
\label{sec1}
From Eqs.~\eqref{eq20} and \eqref{eq28}, 
we can observe that, for the two terms in the numerical flux, 
since the central flux term can be applied 
with the symmetrical conservative metric method,
free-stream preserving can be achieved by canceling 
the dissipative term when the flow is uniform.
In previous work,
this is done either by replacing the transformed conservative variables 
in the difference operator of the dissipative term 
with the original ones and simply neglecting the effect of grid Jacobian~\cite{vinokur2002extension, li2017further}, 
or freeing the metric terms at the point $i+1/2$ to construct the upwind flux~\cite{nonomura2015new}.
In this work, we split the difference operator 
into several local differences involving only two successive grid points.
Since there is a local grid Jacobian shared by each of these differences,
its effect to the overall difference operator is largely preserved.    

\subsection{Free-stream preserving linear-upwind scheme}
\label{subsec1}
The dissipation term ${\bm{\tilde{F}}}_{i+\frac{1}{2}}^{D}$ 
of the linear-upwind numerical flux in Eq. \eqref{eq20} 
can be rewritten into a local difference form as follows:
\begin{equation}
\begin{split}
&{\bm{\tilde{F}}}_{i+\frac{1}{2}}^{D}=
\frac{1}{60}\sum_{s}\bm{R}_{i+\frac{1}{2}}^s  {\lambda}^s  \bm{L}_{i+\frac{1}{2}}^s \cdot \left[  \left( \bm{\tilde{U}}_{i-2}-\bm{\tilde{U}}_{i-1}\right) - 4\left( \bm{\tilde{U}}_{i-1}-\bm{\tilde{U}}_{i}\right) \right. \\
& \hspace{8mm}\left.  +6\left( \bm{\tilde{U}}_{i}-\bm{\tilde{U}}_{i+1}\right)  -4\left( \bm{\tilde{U}}_{i+1}-\bm{\tilde{U}}_{i+2}\right)  
+\left( \bm{\tilde{U}}_{i+2}-\bm{\tilde{U}}_{i+3} \right)\right].
\end{split} 
\label{eq21}
\end{equation}
Then, we can modify Eq. \eqref{eq21} into a free-stream preserving formulation
\begin{equation}
\begin{split}
& {\bm{\tilde{F}}}_{i+\frac{1}{2}}^{D}=
\frac{1}{60}\sum_{s}\bm{R}_{i+\frac{1}{2}}^s  {\lambda}^s  \bm{L}_{i+\frac{1}{2}}^s \cdot 
\left[ \left( \bm{U}_{i-2}-\bm{U}_{i-1}\right)\left(\frac{1}{J}\right)_{i-\frac{3}{2}} 
- 4\left( \bm{U}_{i-1}-\bm{U}_{i}\right) \left(\frac{1}{J}\right)_{i-\frac{1}{2}}\right. \\
&\hspace{8mm} \left. +6\left( \bm{U}_{i}-\bm{U}_{i+1}\right)\left(\frac{1}{J}\right)_{i+\frac{1}{2}}  
-4\left( \bm{U}_{i+1}-\bm{U}_{i+2}\right)\left(\frac{1}{J}\right)_{i+\frac{3}{2}}  
+\left( \bm{U}_{i+2}-\bm{U}_{i+3} \right)\left(\frac{1}{J}\right)_{i+\frac{5}{2}}\right] ,
\end{split} 
\label{eq22}
\end{equation}
by introducing local averaged grid Jacobian 
$ \left(\frac{1}{J}\right)_{i-\frac{3}{2}} $, $ \left(\frac{1}{J}\right)_{i-\frac{1}{2}} $, 
$ \left(\frac{1}{J}\right)_{i+\frac{1}{2}} $, $ \left(\frac{1}{J}\right)_{i+\frac{3}{2}} $ 
and$ \left(\frac{1}{J}\right)_{i+\frac{5}{2}} $, 
which are evaluated by the 6th-order central scheme. 
Taking $ \left(\frac{1}{J}\right)_{i+\frac{1}{2}} $ as an example,
\begin{equation}
\begin{split}
\left(\frac{1}{J}\right)_{i+\frac{1}{2}}=\frac{1}{60}\left[ \left(\frac{1}{J}\right)_{i-2}-8 \left(\frac{1}{J}\right)_{i-1}
+37 \left(\frac{1}{J}\right)_{i}+37\left(\frac{1}{J}\right)_{i+1}
-8 \left(\frac{1}{J}\right)_{i+2}+ \left(\frac{1}{J}\right)_{i+3}\right] .
\end{split} 
\label{eq23}
\end{equation}
In order to increase the numerical accuracy as far as possible, 
the symmetrical conservative form of the metric terms and Jacobian 
in Eqs.~\eqref{metrics2} and~\eqref{eq13} are employed here.
Note that, this technique can be applied to general explicit linear-upwind schemes. 
Taking a linear-upwind scheme with $6$ points stencil as an example,
\begin{equation}
\begin{split}
&{\bm{\tilde{F}}}_{i+\frac{1}{2}}=\sum\limits_{k=1}^{3} a_{k} \left( \bm{\tilde{F}}_{i+k}+\bm{\tilde{F}}_{i-k+1}\right) + \sum_{s}\bm{R}_{i+\frac{1}{2}}^s  {\lambda}^s  \bm{L}_{i+\frac{1}{2}}^s \cdot \sum\limits_{k=1}^{3} b_{k}\left( \bm{\tilde{U}}_{i+k}-\bm{\tilde{U}}_{i-k+1}\right), \\
\end{split} 
\label{eq200}
\end{equation}
where $a_{k}, b_{k}$ are the linear coefficients and $2\left( a_{1}+a_{2}+a_{3}\right) =1$.
The  dissipative part can be rewritten as
\begin{equation}
\begin{split}
&{\bm{\tilde{F}}}_{i+\frac{1}{2}}^{D}=-\sum_{s}\bm{R}_{i+\frac{1}{2}}^s  {\lambda}^s  \bm{L}_{i+\frac{1}{2}}^s \cdot \left[  b_{3}\left( \bm{\tilde{U}}_{i-2}-\bm{\tilde{U}}_{i-1}\right) +\left(b_{2}+b_{3} \right) \left( \bm{\tilde{U}}_{i-1}-\bm{\tilde{U}}_{i}\right) \right.\\
& \left. +\left( b_{1}+b_{2}+b_{3}\right) \left( \bm{\tilde{U}}_{i}-\bm{\tilde{U}}_{i+1}\right)  +\left( b_{2}+b_{3}\right) \left( \bm{\tilde{U}}_{i+1}-\bm{\tilde{U}}_{i+2}\right) + b_{3} \left( \bm{\tilde{U}}_{i+2}-\bm{\tilde{U}}_{i+3}\right) \right],
\end{split} 
\label{eq202}
\end{equation}
which gives a local difference form. 
Then, the same treatment as Eq.~\eqref{eq22} can be applied.

\subsection{Free-stream preserving WENO scheme}
\label{subsec1}
Unlike the treatment of the dissipative term as in  Ref.~\cite{nonomura2015new}, 
we rewrite Eq. \eqref{eq29}, say $\widehat{f}_{i,1}^{s,+}$, 
into a local difference form
\begin{equation}
\begin{split}
&\widehat{f}_{i,1}^{s,+}=f_{i+1}^{s,+}-3f_{i}^{s,+}+3f_{i-1}^{s,+}-f_{i-2}^{s,+}, \\
&=\frac{1}{2}\bm{L}_{i+\frac{1}{2}}^{s}\cdot\left[ \left( \bm{\tilde{F}}_{i+1}-\bm{\tilde{F}}_{i}\right) 
- 2\left( \bm{\tilde{F}}_{i}-\bm{\tilde{F}}_{i-1}\right)  +\left( \bm{\tilde{F}}_{i-1} -\bm{\tilde{F}}_{i-2} \right)\right] \\
&+\frac{1}{2}\lambda^s\bm{L}_{i+\frac{1}{2}}^{s}\cdot\left[ \left( \bm{\tilde{U}}_{i+1}-\bm{\tilde{U}}_{i}\right) 
- 2\left( \bm{\tilde{U}}_{i}-\bm{\tilde{U}}_{i-1}\right)  +\left( \bm{\tilde{U}}_{i-1} -\bm{\tilde{U}}_{i-2} \right)\right]. \\
\end{split} 
\label{eq30}
\end{equation}
Then, Eq.~\eqref{eq30} can be modified for free-stream preserving, similar to Eq.~\eqref{eq22}, as
\begin{equation}
\begin{split}
&\widehat{f}_{i,1}^{p,+}=\frac{\bm{L}_{i+\frac{1}{2}}^{s}}{2}\cdot \left[ \left( \bm{F}_{i+1}-\bm{F}_{i}\right)\left( \frac{\xi_{x}}{J}\right) _{i+\frac{1}{2}} - 2\left( \bm{F}_{i}-\bm{F}_{i-1}\right)\left( \frac{\xi_{x}}{J}\right) _{i-\frac{1}{2}}  +\left( \bm{F}_{i-1} -\bm{F}_{i-2} \right)\left( \frac{\xi_{x}}{J}\right) _{i-\frac{3}{2}}\right] \\
&+\frac{\bm{L}_{i+\frac{1}{2}}^{s}}{2}\cdot \left[ \left( \bm{G}_{i+1}-\bm{G}_{i}\right)\left( {\frac{\xi_{y}}{J}}\right) _{i+\frac{1}{2}} - 2\left( \bm{G}_{i}-\bm{G}_{i-1}\right)\left( \frac{\xi_{y}}{J}\right) _{i-\frac{1}{2}}  +\left( \bm{G}_{i-1} -\bm{G}_{i-2} \right)\left( \frac{\xi_{y}}{J}\right) _{i-\frac{3}{2}}\right] \\
&+\frac{\bm{L}_{i+\frac{1}{2}}^{s}}{2}\cdot \left[ \left( \bm{H}_{i+1}-\bm{H}_{i}\right)\left( {\frac{\xi_{z}}{J}}\right) _{i+\frac{1}{2}} - 2\left( \bm{H}_{i}-\bm{H}_{i-1}\right)\left( \frac{\xi_{z}}{J}\right) _{i-\frac{1}{2}}  +\left( \bm{H}_{i-1} -\bm{H}_{i-2} \right)\left( \frac{\xi_{z}}{J}\right) _{i-\frac{3}{2}}\right] \\
&+\frac{\lambda^s}{2}\bm{L}_{i+\frac{1}{2}}^{s}\cdot \left[ \left( \bm{U}_{i+1}-\bm{U}_{i}\right)\left( {\frac{1}{J}}\right) _{i+\frac{1}{2}} - 2\left( \bm{U}_{i}-\bm{U}_{i-1}\right)\left( \frac{1}{J}\right) _{i-\frac{1}{2}}  +\left( \bm{U}_{i-1} -\bm{U}_{i-2} \right)\left( \frac{1}{J}\right) _{i-\frac{3}{2}}\right], \\
\end{split} 
\label{eq31}
\end{equation}
where the half-point metrics and Jacobians, say $\left( \frac{\xi_{x}}{J}\right) _{i+\frac{1}{2}}$, 
$\left( \frac{\xi_{y}}{J}\right) _{i+\frac{1}{2}}$, $\left( \frac{\xi_{z}}{J}\right) _{i+\frac{1}{2}}$ 
and $\left( \frac{1}{J}\right) _{i+\frac{1}{2}}$ 
are all evaluated with a 6th-order central scheme as in Eq. \eqref{eq23}.

Furthermore, in order to achieve free-stream preserving for the smooth indicators as in Eq. \eqref{eq27},
they are rewritten into a local difference formulation too, say $\beta_0^{+}$ as
\begin{equation}
\beta_0^{+} =\frac{1}{4}\left[\left( f_{i-2}^{s,+}-f_{i-1}^{s,+}\right)-3\left( f_{i-1}^{s,+}-f_{i}^{s,+}\right)  \right]^2
+\frac{13}{12}\left[\left( f_{i-2}^{s,+}-f_{i-1}^{s,+}\right) -\left( f_{i-1}^{s,+}-f_{i}^{s,+}\right)  \right]^2.
\label{eq027}
\end{equation}
Then, the same treatment as Eq.~\eqref{eq31} can be applied.

\subsection{Free-stream preserving hybrid WENO scheme}
\label{subsec1}
Based on the above free-stream preserving linear-upwind and WENO schemes,
it is easy to introduce hybridization following Hu et al.~\cite{hu2015efficient}
to achieve less numerical dissipation and higher computational efficiency.
Here, the non-dimensional discontinuity detector in the characteristic space is
\begin{equation}
\sigma_s=\left(\frac{\Delta v_{i+\frac{1}{2},s}}{\tilde{\rho}} \right) ^{2}, \\
\label{eq32}
\end{equation}
where
\begin{equation}
\begin{split}
& \Delta v_{i+\frac{1}{2},s}=
\frac{1}{60}\bm{L}_{i+\frac{1}{2}}^s \cdot \left[ \left( \bm{U}_{i-2}-\bm{U}_{i-1}\right)\left( \frac{1}{J}\right) _{i-\frac{3}{2}} - 4\left( \bm{U}_{i-1}-\bm{U}_{i}\right) \left( \frac{1}{J}\right) _{i-\frac{1}{2}}\right. \\
&\left. +6\left( \bm{U}_{i}-\bm{U}_{i+1}\right)\left( \frac{1}{J}\right) _{i+\frac{1}{2}}  -4\left( \bm{U}_{i+1}-\bm{U}_{i+2}\right)\left( \frac{1}{J}\right) _{i+\frac{3}{2}}  +\left( \bm{U}_{i+2}-\bm{U}_{i+3} \right)\left( \frac{1}{J}\right) _{i+\frac{5}{2}}\right] ,
\end{split} 
\label{eq33}
\end{equation}
and $\tilde{\rho}$ is the Roe-average density of $\bm{A}_{i+\frac{1}{2}}$.
The threshold is given as
\begin{equation}
\varepsilon=C\left( \frac{\Delta \xi}{L_{\xi}}\right) ^{\alpha}, \\
\label{eq34}
\end{equation}
where $C$ is a positive constant, $\alpha$ is a positive integer and the relation between $L_{\xi}$ and $\Delta \xi$ is defined in Eq.\eqref{eqdelta-xi}. 
Hence, the numerical flux of the hybrid scheme in characteristic space can be switched between that of the linear-upwind  $\tilde{F}_{i+\frac{1}{2}}^{UPS}$ 
and WENO $\tilde{F}_{i+\frac{1}{2}}^{WENO}$ schemes as  
\begin{equation}
\tilde{F}_{i+\frac{1}{2}}=\sigma_{i+\frac{1}{2}} \tilde{F}_{i+\frac{1}{2}}^{UPS}+\left(1-\sigma_{i+\frac{1}{2}} \right)\tilde{F}_{i+\frac{1}{2}}^{WENO},  \\
\label{eq35}
\end{equation}
where $\sigma_{i+\frac{1}{2}}$ which equals to zero when $\sigma_s>\varepsilon$ otherwise one. 

\section{Numerical tests}
\label{sec1}
To demonstrate the effectiveness of the proposed method, 
several problems including free-stream,  isentropic vortex convection, double Mach reflection 
and flow pass a cylinder are computed on various non-uniform grids. 
The local Lax-Friedrichs flux splitting is used for the free-stream and vortex problems and
the global Lax-Friedrichs flux splitting for the double Mach reflection and flow past cylinder problems.
The third order TVD Runge-Kutta scheme is utilized for time integration. 
For hybrid WENO scheme, the parameters are chosen as $C=100$ and $\alpha=3$. 
In the following, while "UPW5" denotes the standard 5th-order linear upwind scheme and 
"WENO" for the classic 5th-order WENO scheme,  
"UPW5-UFP", "WENO-UFP"  and "WENO-HUFP" denote 
the linear upwind scheme, 
the WENO scheme and the hybrid WENO scheme proposed in this paper, respectively, 
and "Exact" denotes the exact solution.

\subsection{Free-stream}
\label{subsec1}
The free-stream is tested on a three dimensional wavy grid 
and a random grid, i.e. a randomly disturbed Cartesian grid, respectively. 
The initial condition of an ideal gas is given as
\begin{equation}
u=u_\infty, \quad v=0, \quad w=0, \quad \rho=\rho_\infty, \quad p=p_\infty.   \\
\label{eq36}
\end{equation}
In this test, we set $u_\infty=0.5$, $\rho_\infty=1$ and $p_\infty=1/\gamma$,
where $\gamma$ is the specific heat ratio.
The Mach number is $0.5$, the same as in Ref.~\cite{nonomura2015new}. 
The states at boundaries are set to the same as those of the initial condition. 
The time-step size is set to $0.1$ and the results are examined after $100$ time steps.

\begin{figure}\centering
	\subfigure [Wavy grid]{\includegraphics[width=0.42\textwidth]{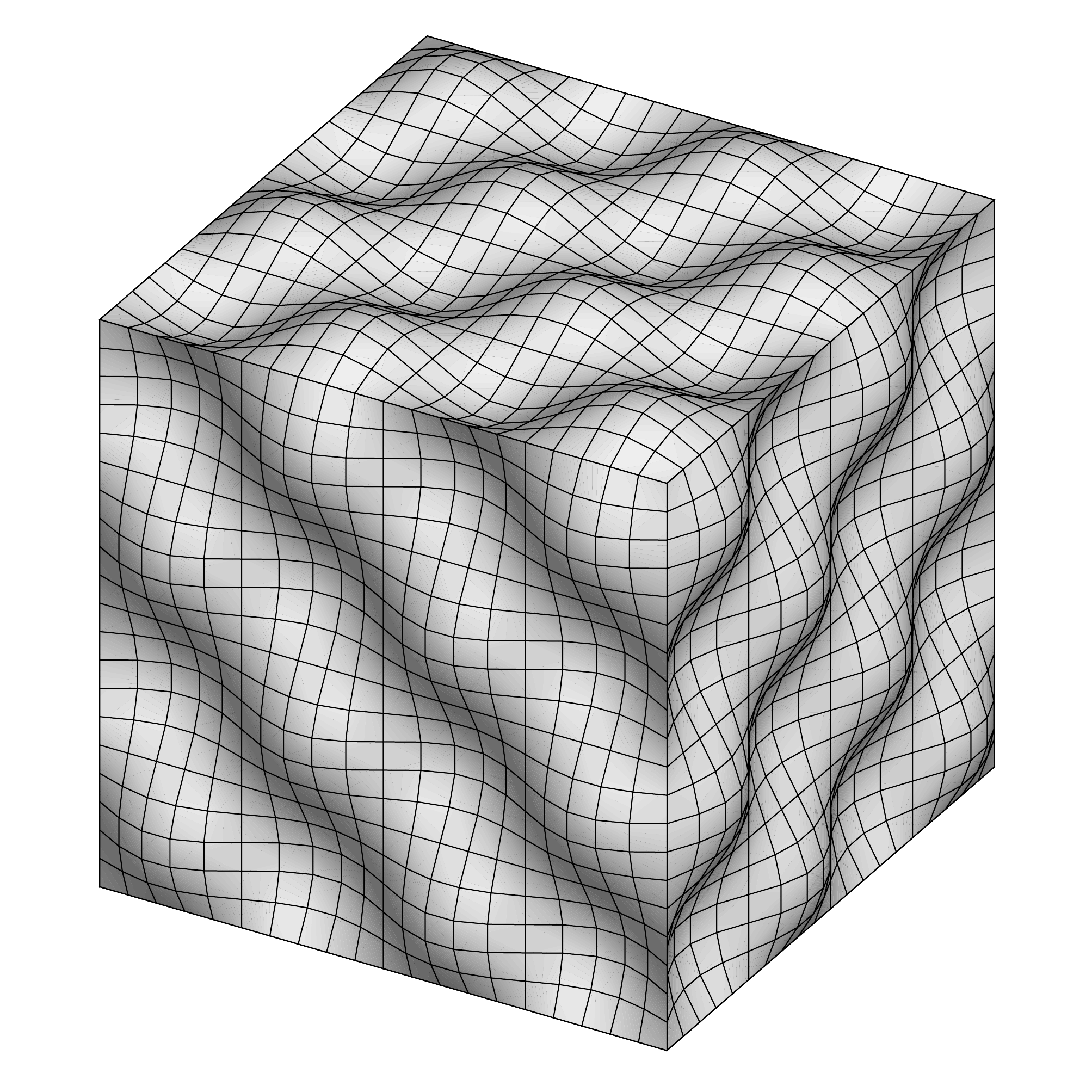}}
	\subfigure [ Random grid]{\includegraphics[width=0.42\textwidth]{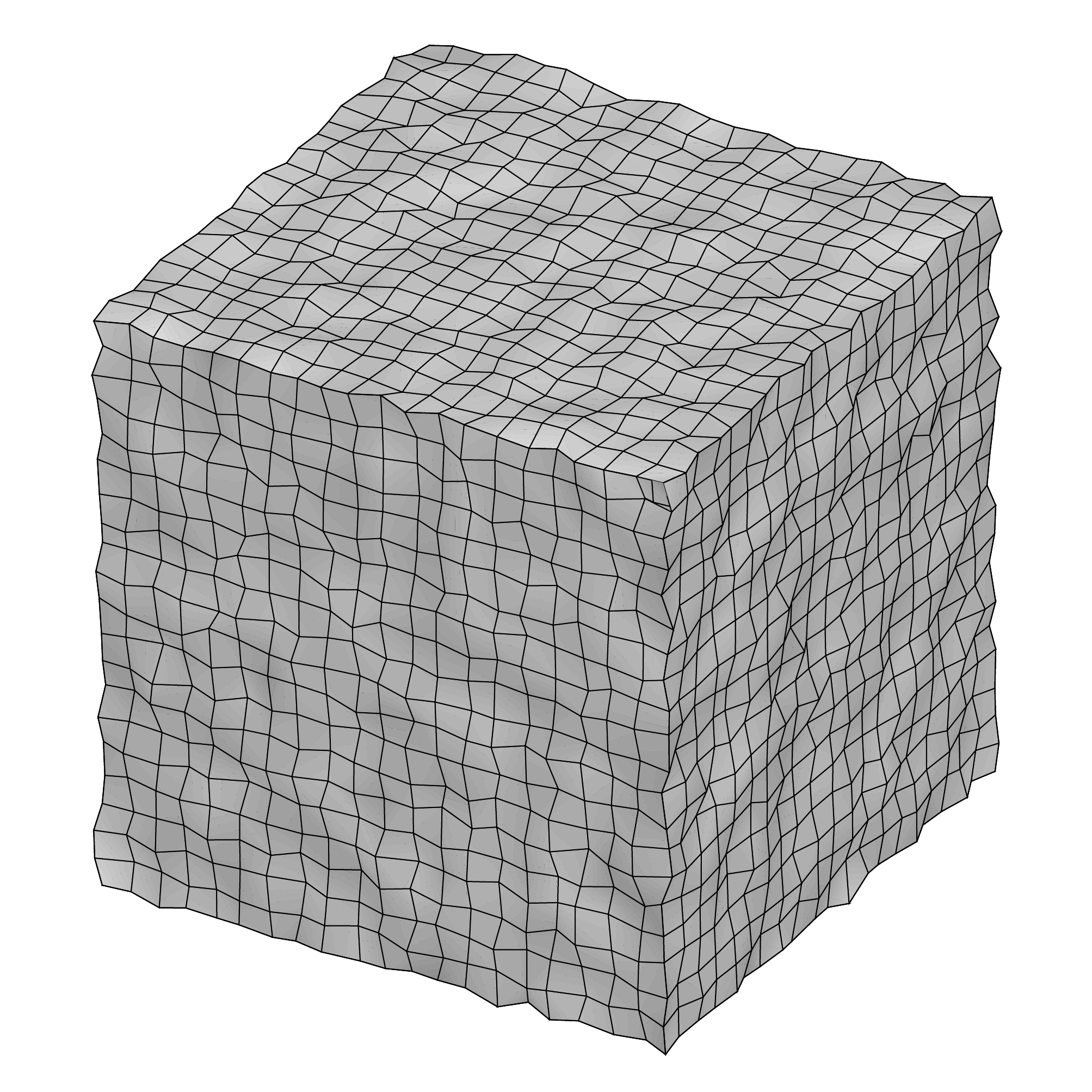}}
	\caption{ The three-dimensional wavy and random grids for the free-stream problem.  }
	\label{3d-grid}
\end{figure}

First, we test the free-stream preservation property on the wavy grid, 
as shown in Fig.~\ref{3d-grid}(a), defined in the domain 
$\left[ -2,2\right] \times\left[ -2,2\right]\times \left[ -2,2\right]$ 
by
\begin{equation}
\begin{split}
&x_{i,j,k}=x_{min}+\Delta x_0\left[\left( i-1\right)+A_x sin\frac{n_{xy}\pi\left(j-1 \right)\Delta y_0 }{L_y} sin\frac{n_{xz}\pi\left(k-1 \right)\Delta z_0 }{L_z} \right] , \\
&y_{i,j,k}=y_{min}+\Delta y_0\left[\left( j-1\right)+A_y sin\frac{n_{yz}\pi\left(k-1 \right)\Delta z_0 }{L_z} sin\frac{n_{yx}\pi\left(i-1 \right)\Delta x_0 }{L_x} \right] , \\
&z_{i,j,k}=z_{min}+\Delta z_0\left[\left( k-1\right)+A_z sin\frac{n_{zx}\pi\left(i-1 \right)\Delta x_0 }{L_x} sin\frac{n_{zy}\pi\left(j-1 \right)\Delta y_0 }{L_y} \right] , \\
\end{split} 
\label{eq37}
\end{equation}
where $L_x=L_y=L_z=4$, $A_x=A_y=A_z=1$, $n_{xy}=n_{xz}=n_{yz}=n_{yx}=n_{zx}=n_{zy}=4$, 
and $x_{min}=-L_x/2$, $y_{min}=-L_y/2$, $z_{min}=-L_z/2$. 
The grid resolution is set to $21\times 21$. 
The $L_2$ errors of the velocity components $v$ and $w$ 
of the flow field are shown in Table~\ref{3d-wavy-result}. 
One can find that the errors of UPW5-UFP, WENO-UFP and WENO-HUFP are 
all less than $10^{-15}$, which is close to machine zero. 
However, the results obtained by UPW5 and WENO scheme exhibit much large errors. 
This test demonstrates that the present method is effectively free-stream preserving. 
Note that, the errors of WENO-HUFP are the same with that of UPW5-UFP, 
implying that only the linear-upwind scheme of the hybrid WENO method is switched on 
throughout the entire computation.
\begin{table}
	\scriptsize
	\centering
	\caption{$L_2$ errors of $v$ and $w$ components in the free-stream problem on the wavy grid}
	\begin{tabularx}{13.5cm}{@{\extracolsep{\fill}}llr}
		\hline
		Method & $v-$component &  $w-$component \\
		\hline
		UPW5& $1.56\times 10^{-3}$	& $2.48\times 10^{-3}$ \\
		WENO & $9.25\times 10^{-3}$ & $1.03\times 10^{-2}$ \\
		UPW5-UFP	& $6.91\times 10^{-16}$	& $5.70\times 10^{-16}$ \\
		WENO-UFP	& $6.99\times 10^{-16}$	& $6.86\times 10^{-16}$ \\
		WENO-HUFP	& $6.91\times 10^{-16}$	& $5.70\times 10^{-16}$ \\
		\hline
	\end{tabularx}
	\label{3d-wavy-result}
\end{table}

Similar to the wavy grid, as shown in Fig. \ref{3d-grid}(b), 
the random grid has the same domain and grid resolution,
but generated by
\begin{equation}
\begin{split}
&x_{i,j,k}=x_{min}+\Delta x_0\left[\left( i-1\right)+A_x \left(2\varphi_x-1 \right)  \right] , \\
&y_{i,j,k}=y_{min}+\Delta y_0\left[\left( j-1\right)+A_y \left(2\varphi_y-1 \right)  \right] , \\
&z_{i,j,k}=z_{min}+\Delta z_0\left[\left( k-1\right)+A_z \left(2\varphi_z-1 \right)  \right] , \\
\end{split} 
\label{eq38}
\end{equation}
where $A_x=A_y=A_z=0.2$ 
are magnitudes of the random disturbances 
and $\varphi_x$, $\varphi_y$ ,$\varphi_z$ 
are random numbers uniformly distributed between $0$ and $1$.
The $L_2$ errors of velocity components $v$ and $w$ of the flow field are 
shown in Table~\ref{3d-random-result}. 
\begin{table}
	\scriptsize
	\centering
	\caption{$L_2$ errors of $v$ and $w$ components in the free-stream problem on the random grid}
	\begin{tabularx}{13.5cm}{@{\extracolsep{\fill}}llr}
		\hline
		Method & $v-$component &  $w-$component \\
		\hline
		UPW5& $4.91\times 10^{-2}$	& $2.94\times 10^{-2}$ \\
		WENO & $1.25\times 10^{-1}$ & $7.81\times 10^{-2}$ \\
		UPW5-UFP	& $6.91\times 10^{-16}$	& $5.31\times 10^{-16}$ \\
		WENO-UFP	& $6.86\times 10^{-16}$	& $6.70\times 10^{-16}$ \\
		WENO-HUFP	& $6.91\times 10^{-16}$	& $5.31\times 10^{-16}$ \\
		\hline
	\end{tabularx}
	\label{3d-random-result}
\end{table}
These results also prove that the present method eliminates geometrically 
induced errors to a large extend and preserves free-stream effectively.

\subsection{Isentropic vortex}
\label{subsec1}
This two-dimensional case, taken from Ref.~\cite{nonomura2015new}, 
is also computed on wavy and random grids to test the vortex preservation property. 
An isentropic vortex centered at $\left( x_c,y_c\right)=\left( 0,0\right)$ 
is superposed to a uniform flow with Mach $0.5$ as the initial condition. 
The perturbations of the isentropic vortex are given by
\begin{equation}
\begin{split}
&\left(\delta u,\delta v \right)=\varepsilon \tau e^{\alpha\left(1-\tau^2 \right) }\left( sin \theta, -cos\theta\right) ,  \\
&\delta T=-\frac{\left(\gamma-1 \right)\varepsilon^2 }{4\alpha \gamma}e^{\alpha\left(1-\tau^2 \right) }, \\
&\delta S=0, \\
\end{split} 
\label{eq39}
\end{equation}
where $\alpha=0.204$, $\varepsilon=0.02$, $\tau=r/r_c$ 
and $r=\left[ \left(x-x_c \right)^2+\left(y-y_c \right)^2 \right]^{1/2} $. 
Here, $r_c=1.0$ is the vortex core length, 
$T=p/\rho$ is the temperature and $S=p/\rho^{\gamma}$ is the entropy. 
The periodic boundary condition is adopted 
and the flow field is examined when the vortex moving back to the original location.
 
The first test is on a wavy grid defined in the domain 
$\left(x,y \right)\in \left[ -10,10\right]\times\left[-10,10 \right]$ by 
\begin{equation}
\begin{split}
&x_{i,j}=x_{min}+\Delta x_0\left[\left( i-1\right)+A_x sin \frac{n_{xy}\pi\left(j-1 \right)\Delta y_0 }{L_y}\right] , \\
&y_{i,j}=y_{min}+\Delta y_0\left[\left( j-1\right)+A_y sin \frac{n_{yx}\pi\left(i-1 \right)\Delta x_0 }{L_x}\right] , \\
\end{split} 
\label{eq40}
\end{equation}
where $L_x=L_y=20$, $x_{min}=-L_x/2$, $y_{min}=-L_y/2$, 
$A_x\times \Delta x_0=0.6$, $A_y\times \Delta y_0=0.6$ and $n_{xy}=n_{yx}=4$. 
In order to evaluate the grid convergence, 
three grids with the resolutions of $21\times 21, 41\times 41, 81\times 81$ are used. 
The time-step sizes $\Delta t$ respect to those grids are select carefully 
as $0.25$, $0.0625$ and $0.0015625$, respectively,
to eliminate the errors induced by time integration.

\begin{figure}\centering
	\subfigure [Exact]{\includegraphics[width=0.32\textwidth]{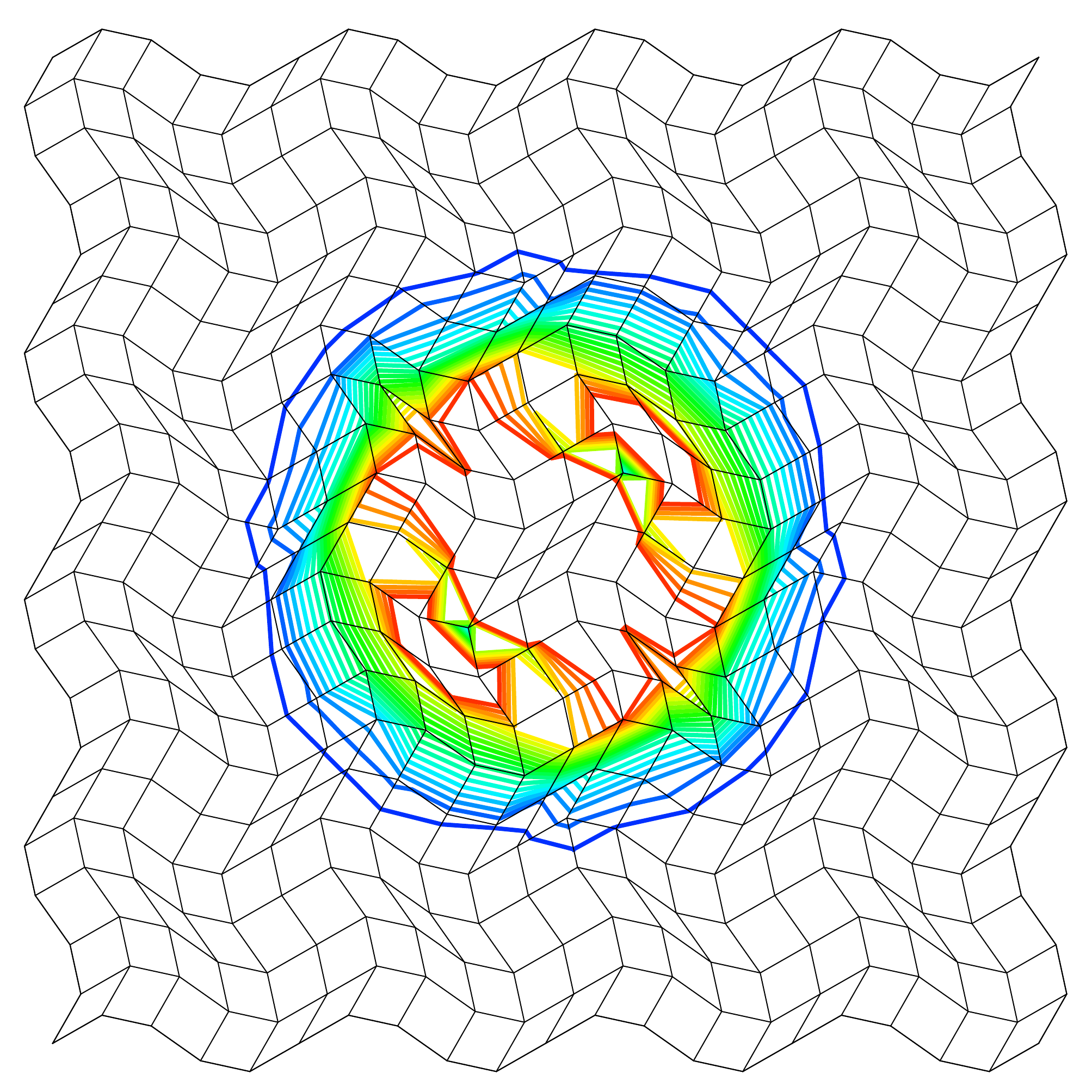}}
	\subfigure [ UPW5]{\includegraphics[width=0.32\textwidth]{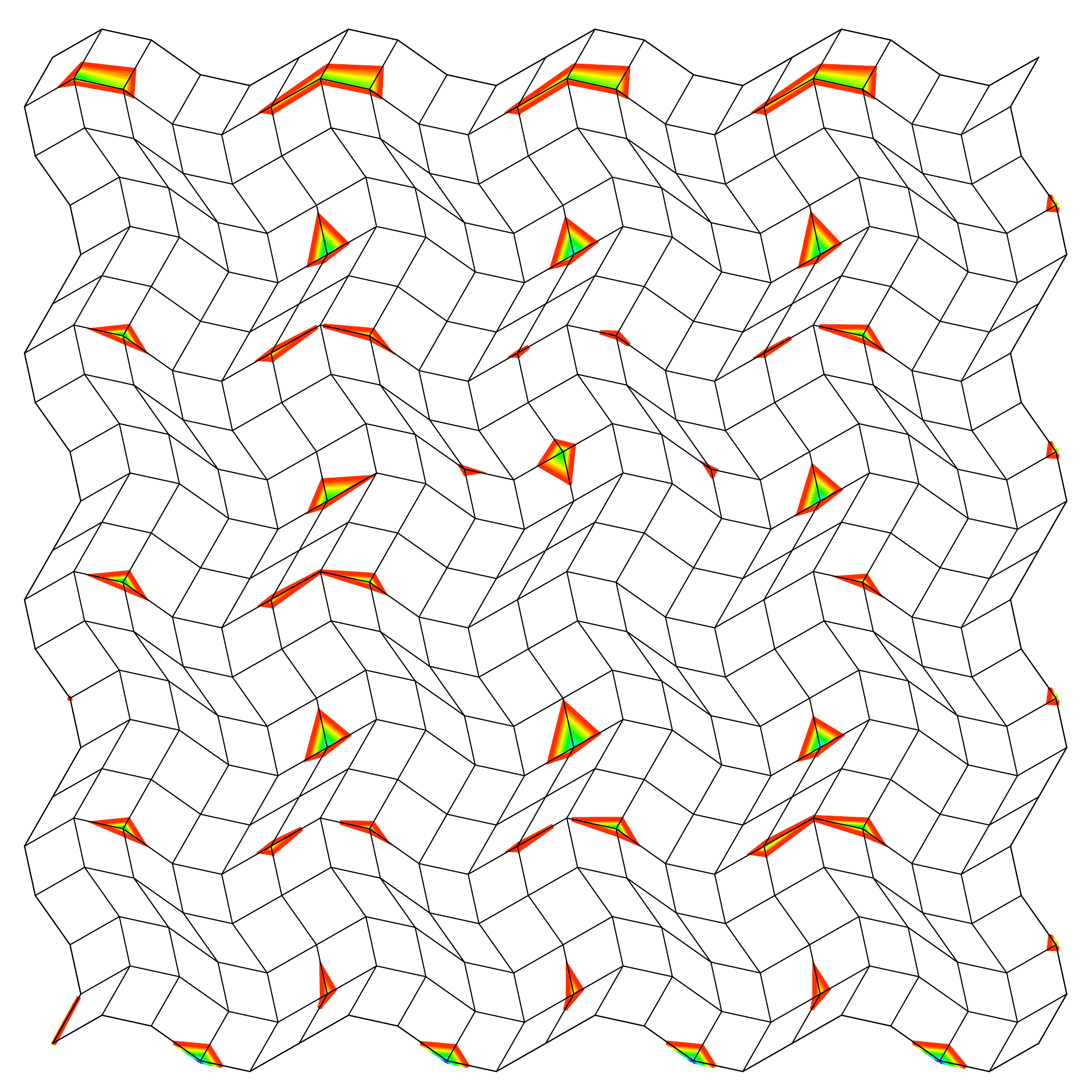}}
	\subfigure [ WENO]{\includegraphics[width=0.32\textwidth]{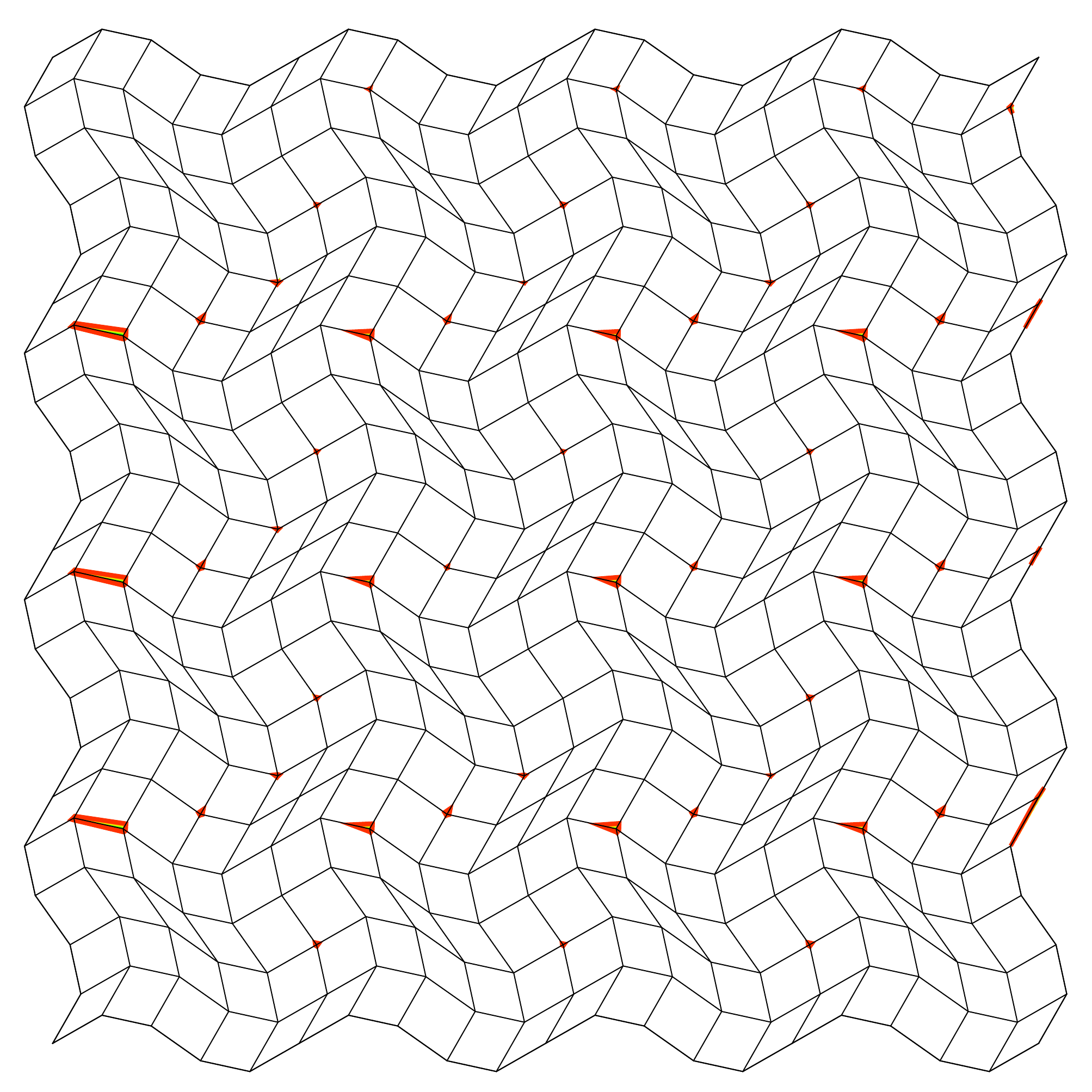}}
	\subfigure [ UPW5-UFP ] {\includegraphics[width=0.32\textwidth]{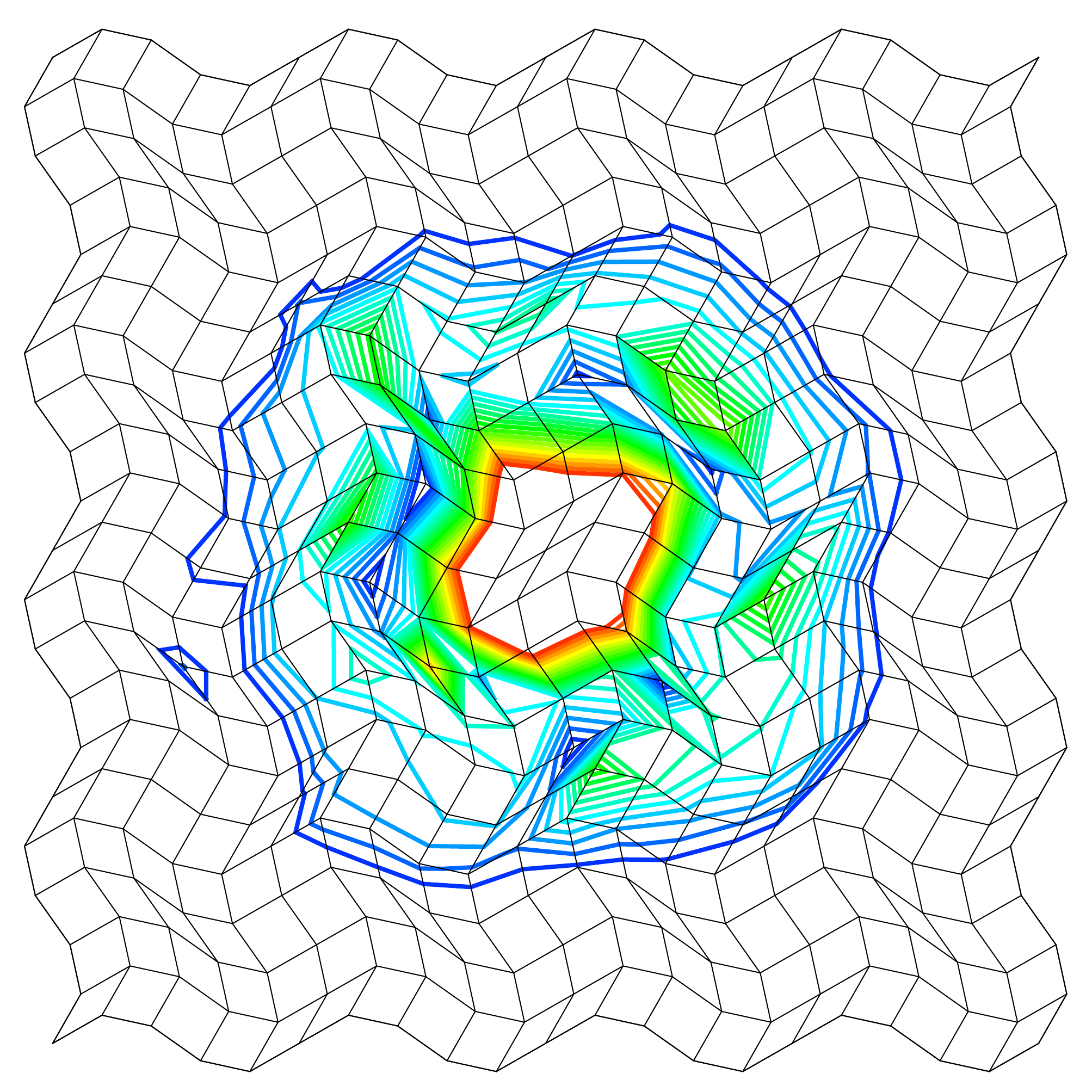}}
	\subfigure [ WENO-UFP ] {\includegraphics[width=0.32\textwidth]{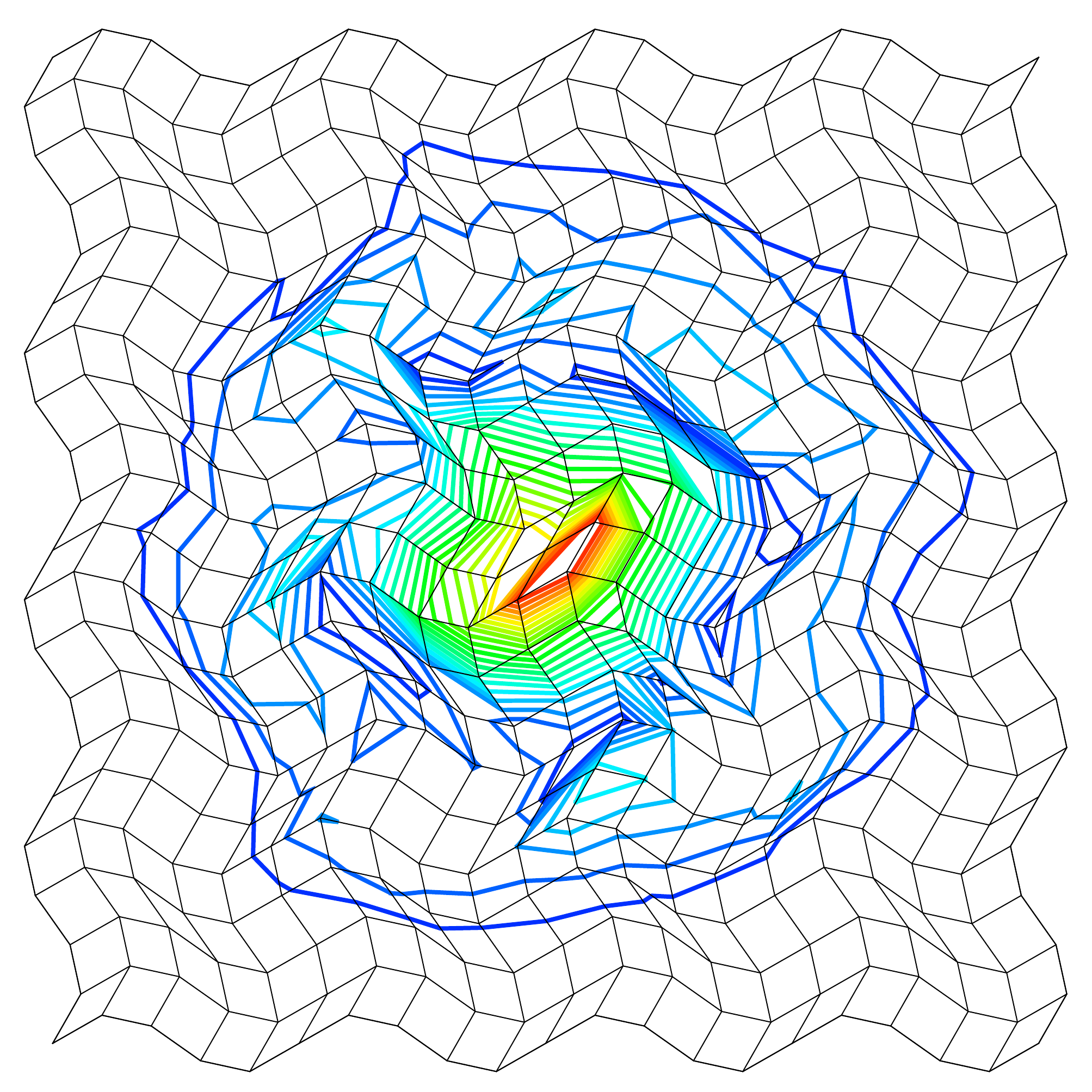}}
	\subfigure [ WENO-HUFP ] {\includegraphics[width=0.32\textwidth]{upw5fp-2d-wavy.pdf}}
	\caption{  21 equally spaced vorticity contours from $0.0$ to $0.006$ of moving vortex on a two dimensional wavy grid.  }
	\label{2d-wavy}
\end{figure}

\begin{figure}
	\centering
	{\includegraphics[width=0.6\textwidth]{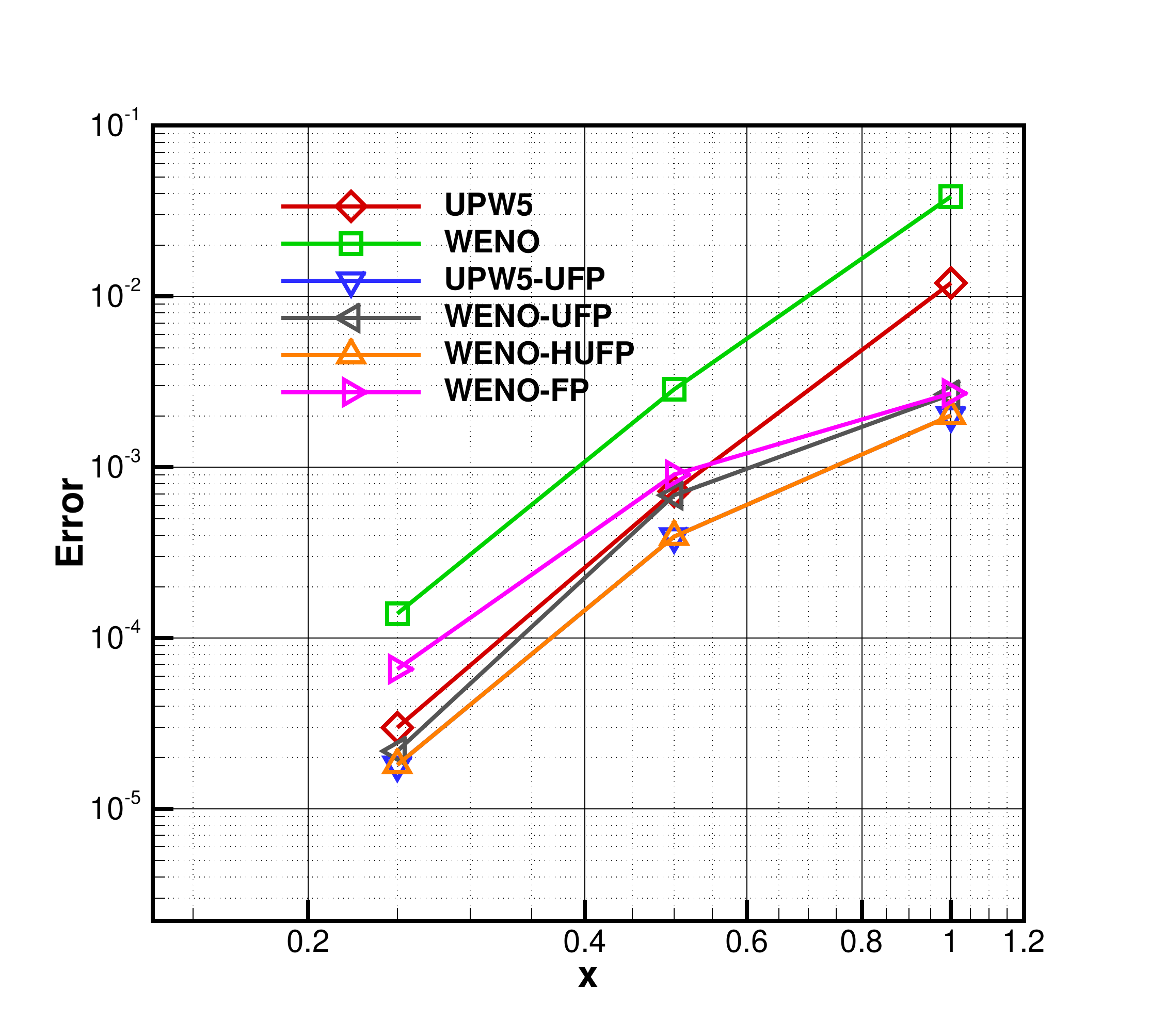}}
	\caption{ Errors of vortex on grids with different density. }
	\label{error-wavy}
\end{figure}

The flow field computed on the $21\times 21$ grid are shown in Fig.~\ref{2d-wavy}. 
It can be observed that UPW and WENO are not able to resolve the moving vortex. 
The errors generated from the wavy grid pollute the entire flow field. 
However, the vortex is resolved well by both UPW5-UFP and WENO-UFP. 
In addition, the flow field obtained by UPW5-UFP is closer to the exact solution 
than that of WENO-UFP. 
Again, the essentially same results obtained by UPW5 and WENO-HUFP
imply that, since there is no discontinuity in the solution, 
only the linear-upwind scheme of the hybrid WENO method is switched on 
throughout the entire computation.
The $L_2$ errors of the $v$ component on wavy grids at three resolutions 
are shown in Table~\ref{2d-wavy-result} and Fig.~\ref{error-wavy}. 
\begin{table}
	\scriptsize
	\centering
	\caption{$L_2$ errors of $v$ component in the vortex problem on different wavy grids}
	\begin{tabularx}{13.5cm}{@{\extracolsep{\fill}}lllr}
		\hline
		Method & Grid number &  Error & Order of accuracy\\
		\hline
		UPW5& $21\times 21$	& $1.20\times 10^{-2}$& $-$ \\
		\quad& $41\times 41$	& $7.23\times 10^{-4}$& $4.05$ \\
		\quad& $81\times 81$	& $2.99\times 10^{-5}$& $4.60$ \\
		WENO & $21\times 21$	& $3.85\times 10^{-2}$& $-$ \\
		\quad& $41\times 41$	& $2.86\times 10^{-3}$& $3.75$ \\
		\quad& $81\times 81$	& $1.39\times 10^{-4}$& $4.36$ \\
		UPW5-UFP & $21\times 21$	& $2.01\times 10^{-3}$& $-$ \\
		\quad& $41\times 41$	& $3.93\times 10^{-4}$& $2.36$ \\
		\quad& $81\times 81$	& $1.81\times 10^{-5}$& $4.44$ \\
		WENO-UFP& $21\times 21$	& $2.70\times 10^{-3}$& $-$ \\
		\quad& $41\times 41$	& $7.24\times 10^{-4}$& $1.90$ \\
		\quad& $81\times 81$	& $2.19\times 10^{-5}$& $5.05$ \\
		WENO-HUFP& $21\times 21$	& $2.01\times 10^{-3}$& $-$ \\
		\quad& $41\times 41$	& $3.93\times 10^{-4}$& $2.36$ \\
		\quad& $81\times 81$	& $1.81\times 10^{-5}$& $4.44$ \\
		\hline
	\end{tabularx}
	\label{2d-wavy-result}
\end{table}
In Fig.~\ref{error-wavy}, WENO-FP denotes the results obtained with 
the method of Nonomura et al. ~\cite{nonomura2015new}.  
These results suggest that the present method works well 
and the errors produced by the UPW5-UFP, WENO-UFP and WENO-HUFP all are lower than WENO-FP,
and have higher convergence rate.

\begin{figure}
	\scriptsize
	\centering
	\subfigure [Exact]{\includegraphics[width=0.32\textwidth]{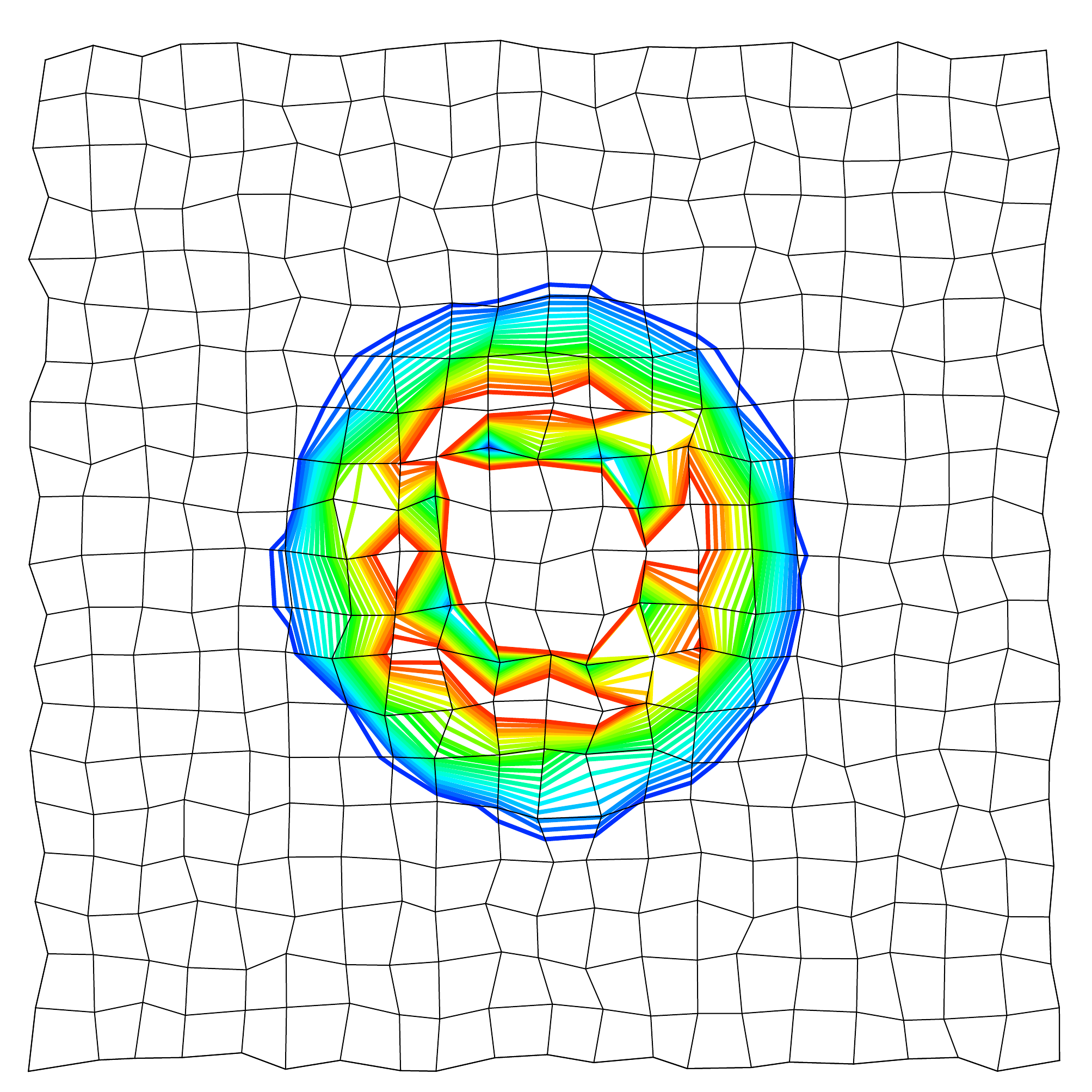}}
	\subfigure [ UPW5]{\includegraphics[width=0.32\textwidth]{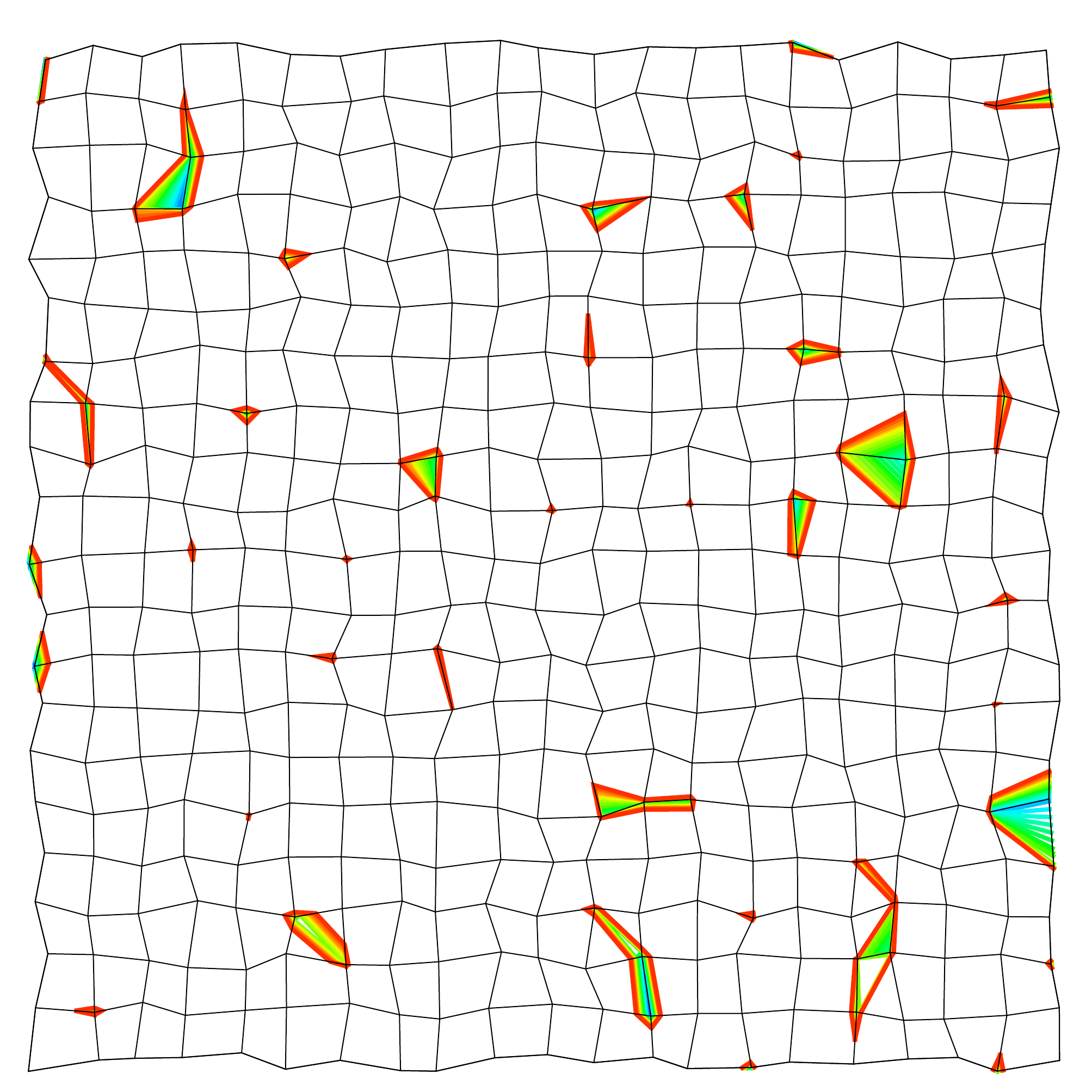}}
	\subfigure [ WENO]{\includegraphics[width=0.32\textwidth]{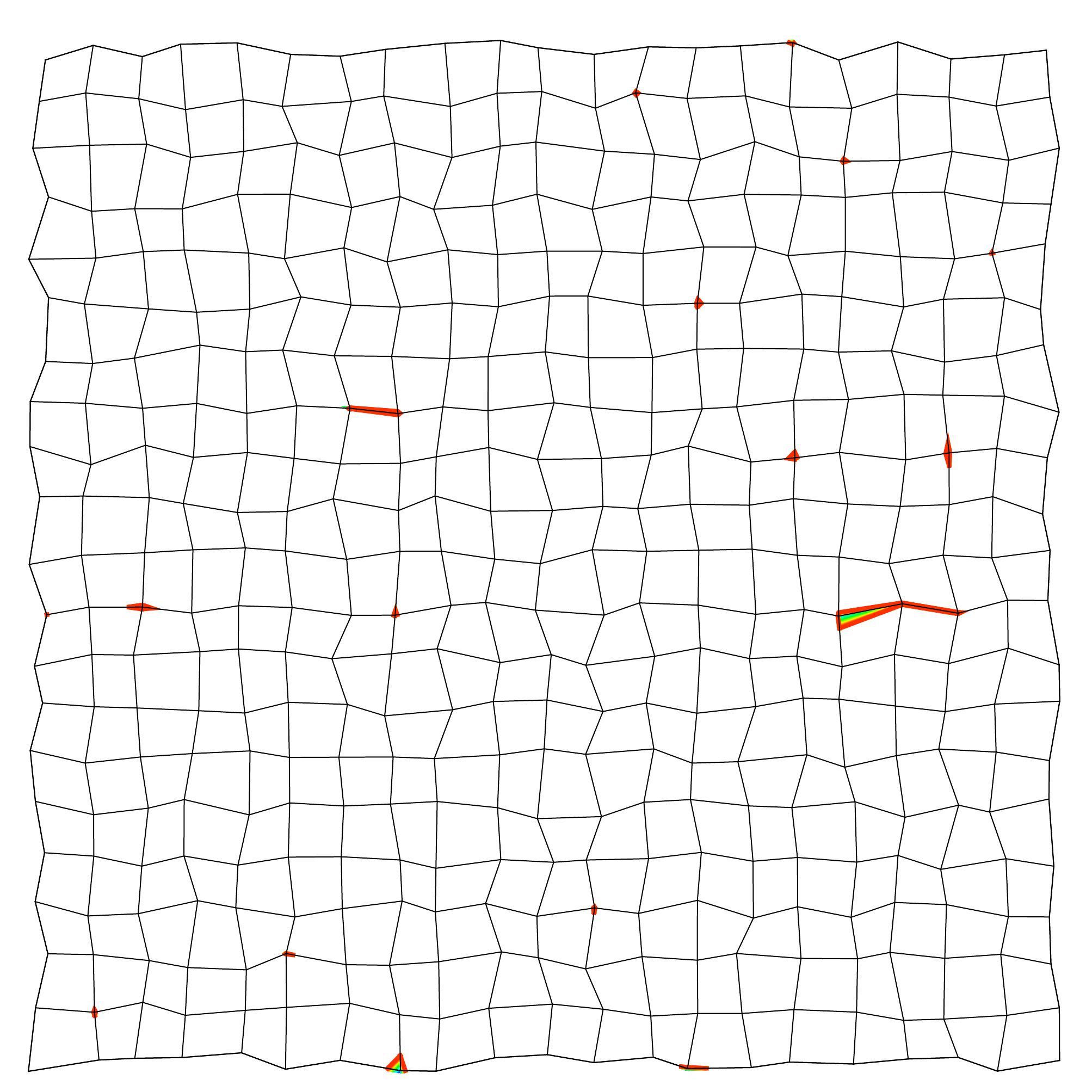}}
	\subfigure [ UPW5-UFP ] {\includegraphics[width=0.32\textwidth]{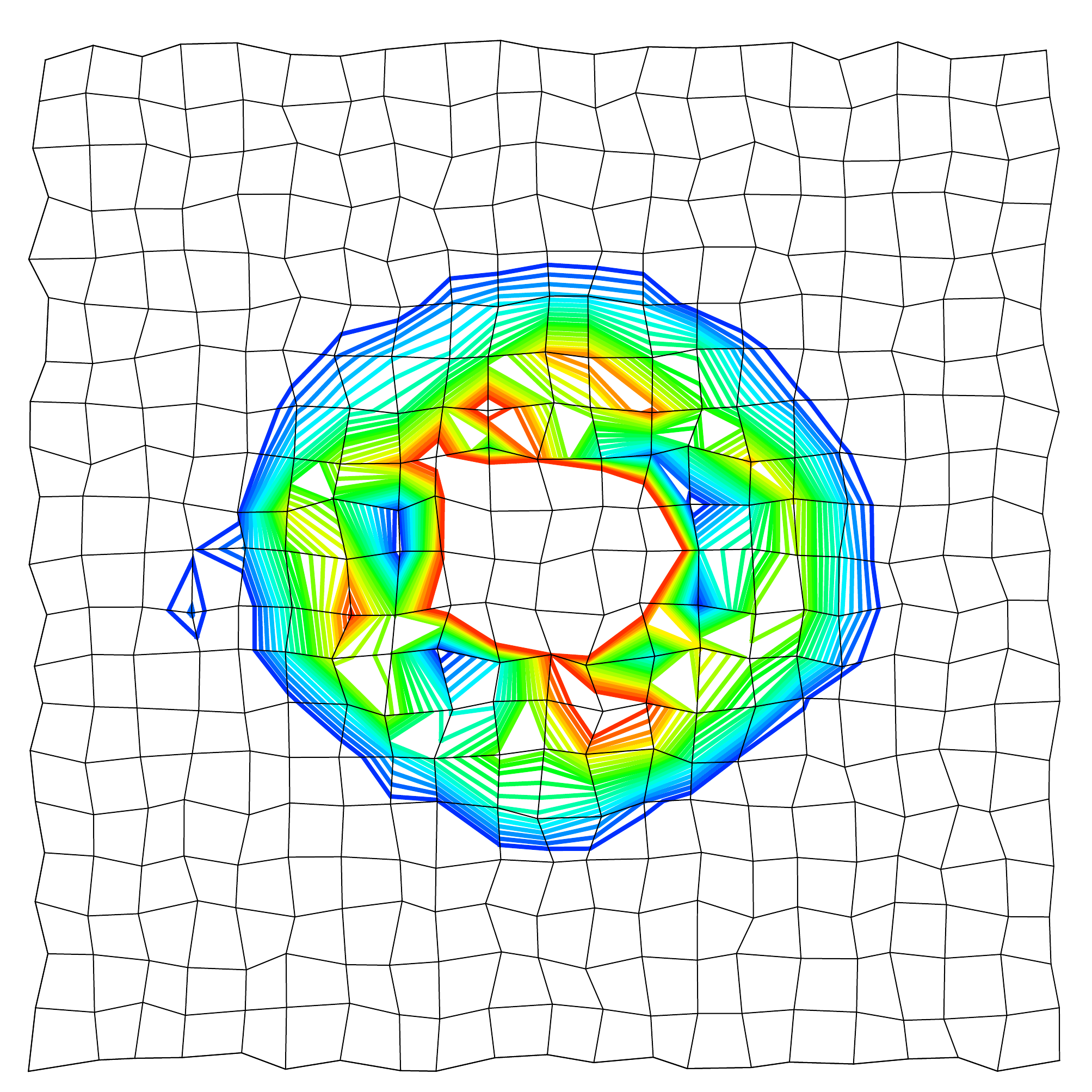}}
	\subfigure [ WENO-UFP ] {\includegraphics[width=0.32\textwidth]{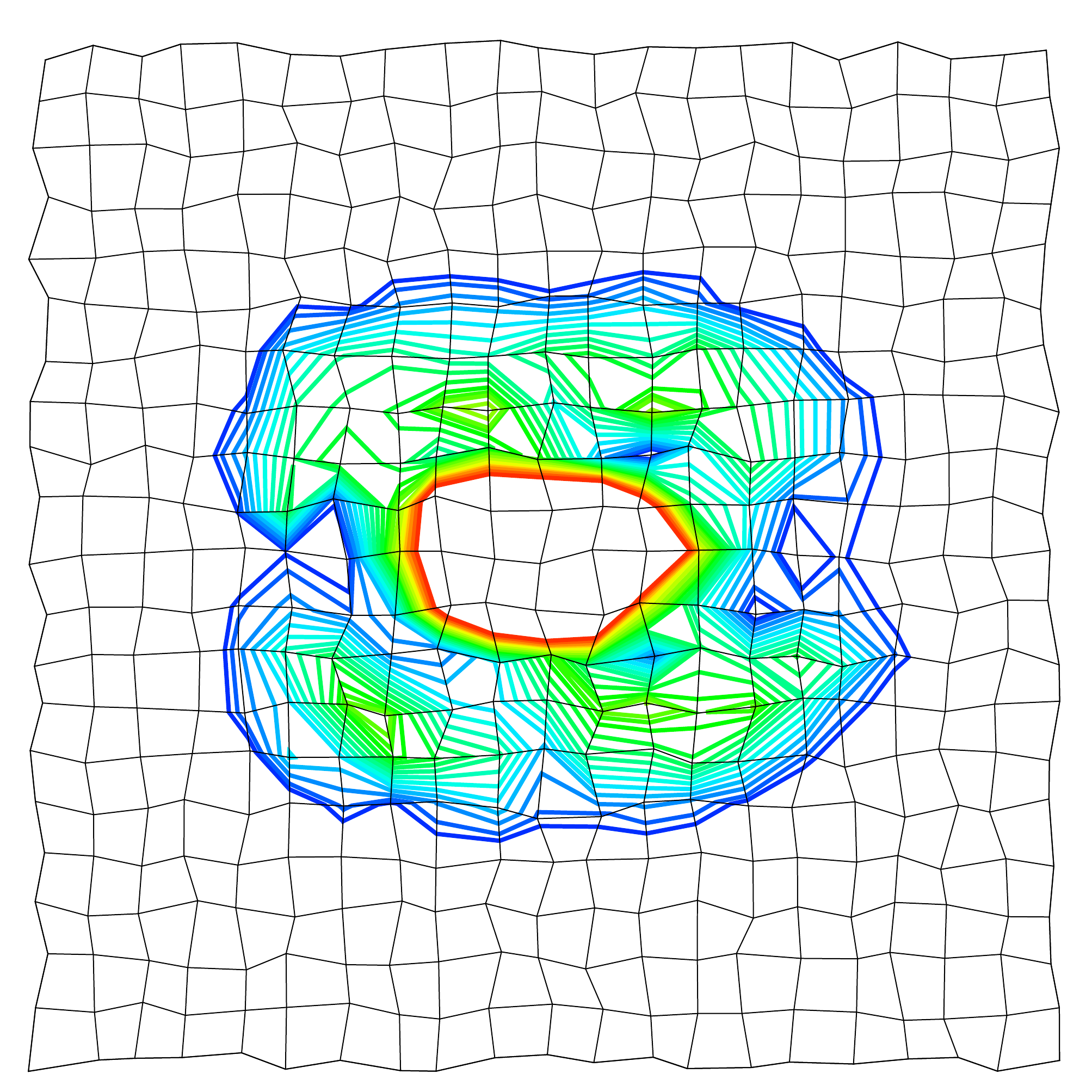}}
	\subfigure [ WENO-HUFP ] {\includegraphics[width=0.32\textwidth]{UPW5-UFP-random.pdf}}
	\caption{  21 equally spaced vorticity contours from $0.0$ to $0.006$ of moving vortex on a two dimensional random grid.  }
	\label{2d-random}
\end{figure}

Then, the vortex is tested on a random grid at the resolution of $21\times 21$ 
and with time-step size of $\Delta t=0.25$. 
The grid points are randomized in a random direction with $20\%$ 
of the original Cartesian grid size. 
The vorticity contours and $L_2$ errors of the $v$ component are shown 
in Fig.~\ref{2d-random} and Table~\ref{2d-random-result}, respectively. 
From these results, it can be observed that the flows computed by UPW5 and WENO produce much larger errors 
and the present schemes preserve the vortex well. 
it is clear shown that the results obtained by the UPW5-UFP and WENO-HUFP 
are better than that of WENO-UFP due to less numerical dissipation.

\begin{table}
	\scriptsize
	\centering
	\caption{$L_2$ errors of $v$ component in the vortex problem on a randomized grid}
	\begin{tabularx}{13.5cm}{@{\extracolsep{\fill}}llr}
		\hline
		Method & Grid number &  Error \\
		\hline
		UPW5& $21\times 21$	& $3.16\times 10^{-2}$\\
		WENO & $21\times 21$	& $4.72\times 10^{-2}$ \\
		UPW5-UFP & $21\times 21$	& $1.34\times 10^{-3}$\\
		WENO-UFP& $21\times 21$	& $2.25\times 10^{-3}$ \\
		WENO-HUFP& $21\times 21$	& $1.34\times 10^{-3}$ \\
		\hline
	\end{tabularx}
	\label{2d-random-result}
\end{table}

\subsection{Double Mach reflection}
\label{subsec1}
\begin{figure}
	\scriptsize
	\centering
	\subfigure [Grid]{\includegraphics[width=0.75\textwidth]{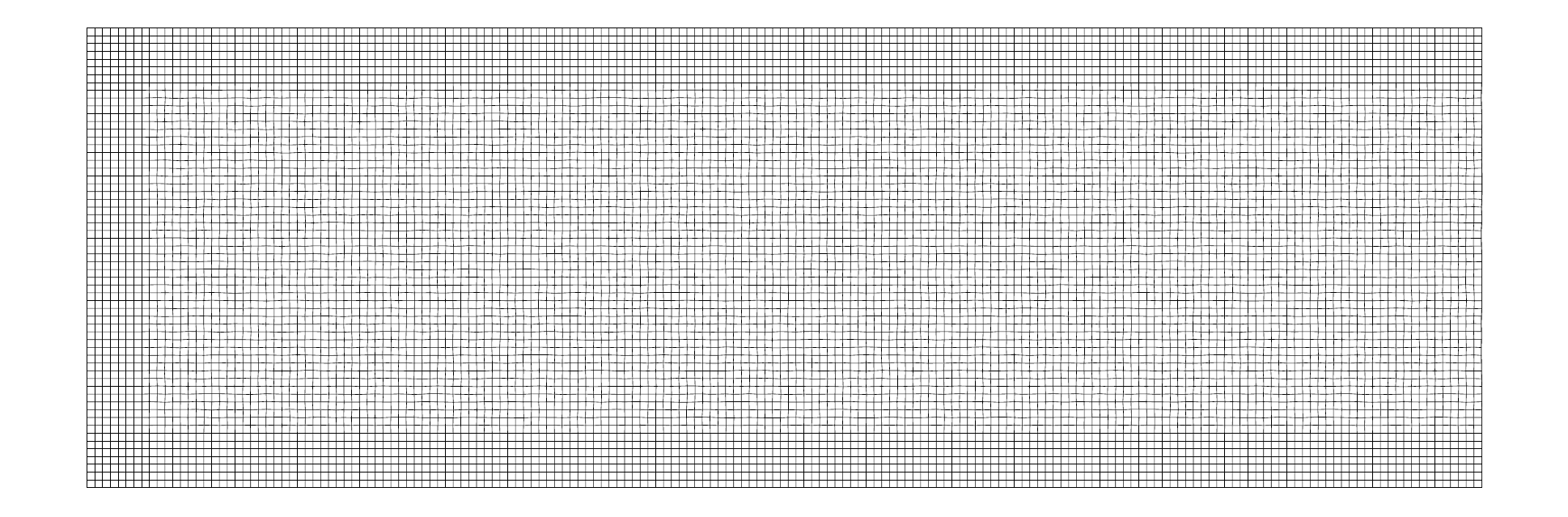}}
	\subfigure [ WENO]{\includegraphics[width=0.75\textwidth]{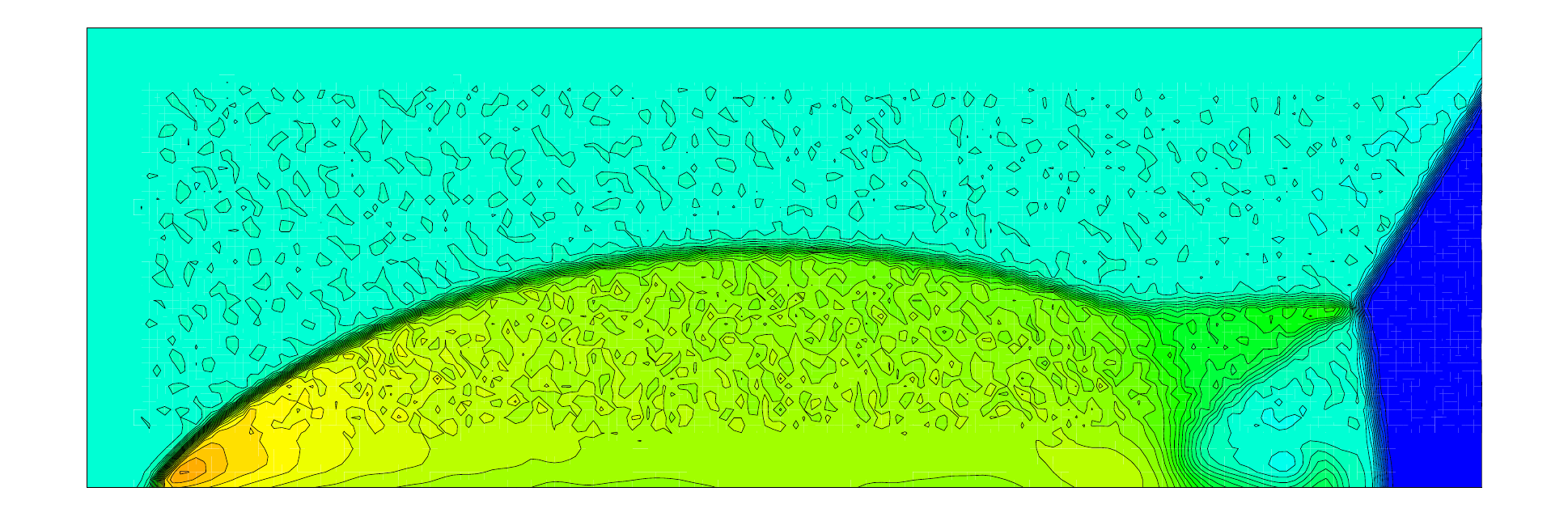}}
	\subfigure [ WENO-UFP]{\includegraphics[width=0.75\textwidth]{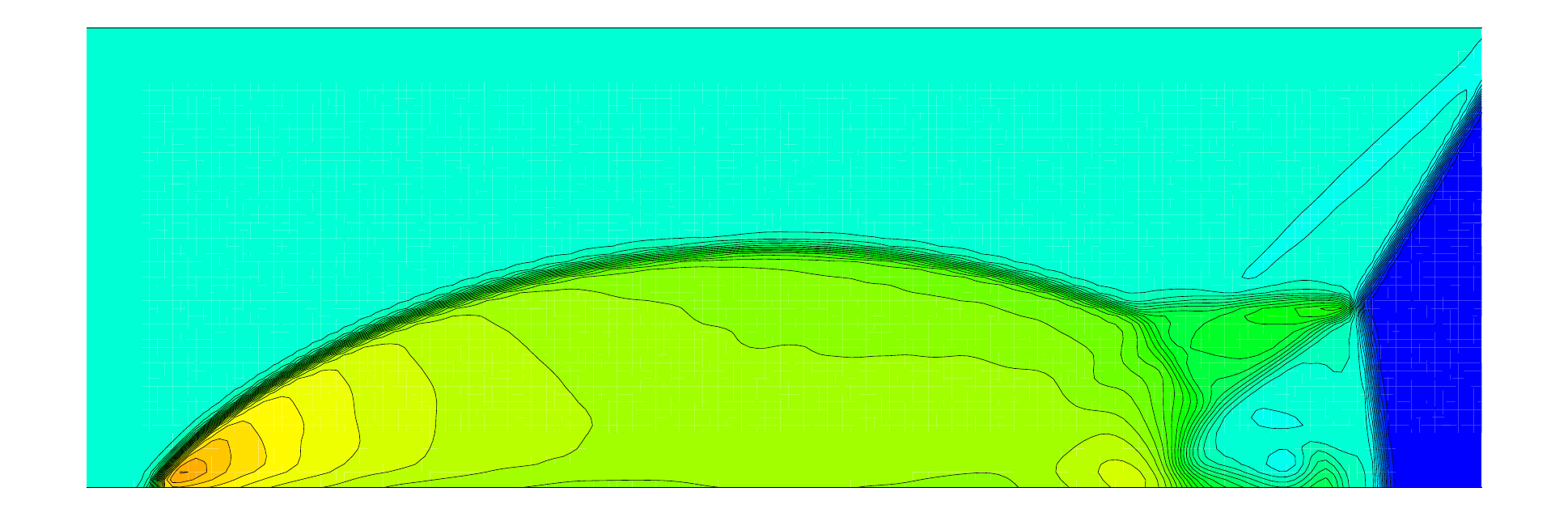}}
	\subfigure [ WENO-HUFP ] {\includegraphics[width=0.75\textwidth]{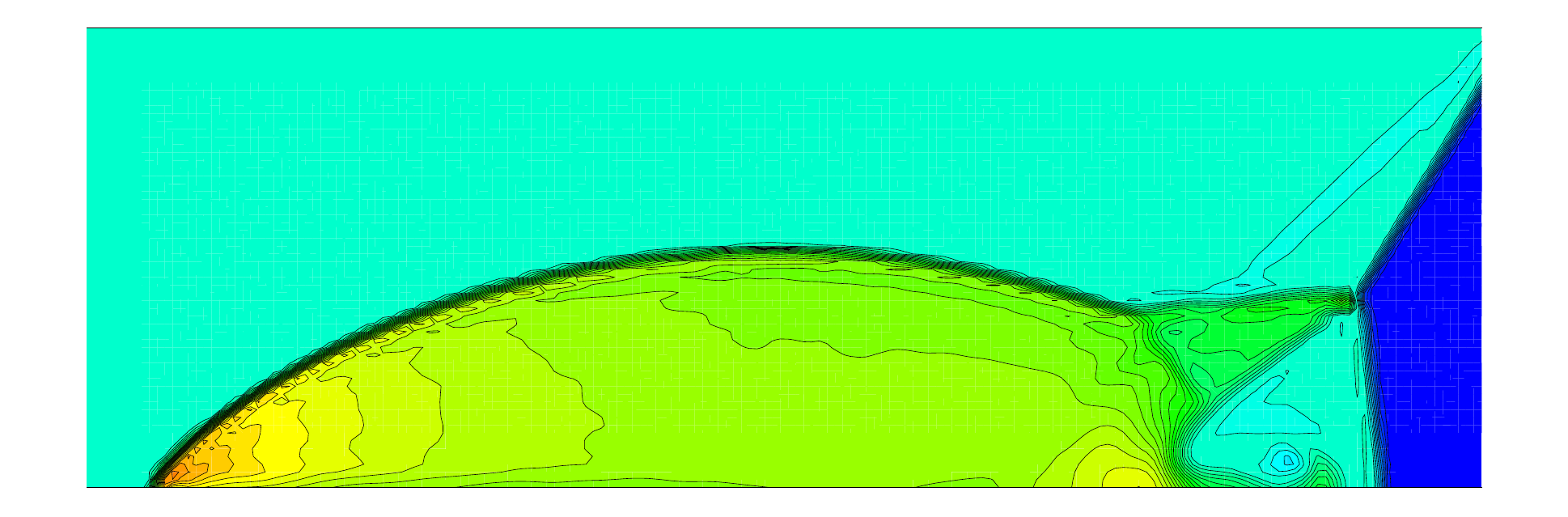}}
	\caption{ 41 equally spaced density contours from 1.92 to 22.59 of 
		double Mach reflection problem on the $240\times 60$ grid.  }
	\label{DM-60}
\end{figure}

The double Mach reflection problem ~\cite{WOODWARD1984115} 
containing strong shock waves is chosen to examine 
the shock-capturing property of the present method. 
In the computational domain $\left(x,y \right)\in \left[0,4 \right] \times\left[ 0,1\right]$, 
the initial conditions are 
\begin{equation}
\left( \rho, u, v, p\right)^T =\left\lbrace 
\begin{array}{rcl}
\left(1.4, 0.0, 0.0, 1.0 \right)^T &    & {x-y \tan\frac{\pi}{6}\geq\frac{1}{6}}\\
\left(8.0, 7.1447, -4.125, 116.5 \right)^T &    & else\\
\end{array}\right..
\label{eq41}
\end{equation}
The computation is conducted up to $t=0.2$ with the CFL number of 0.6. 
Two random grids with $5\%$ randomization at two resolutions of  $240\times 60$ 
and $960\times 240$ are used. 
In order to preserve high accuracy near the boundary, as shown in Fig. \ref{DM-60}(a), 
several points near the edges are left unperturbed. 
Note that $5\%$ randomization is sufficiently sever 
and the errors of inappropriate implementation is significant. 
In Ref.~\cite{nonomura2015new}, only $2\%$ randomization is applied to their test.

\begin{figure}\centering
	\subfigure [WENO]{\includegraphics[width=0.69\textwidth]{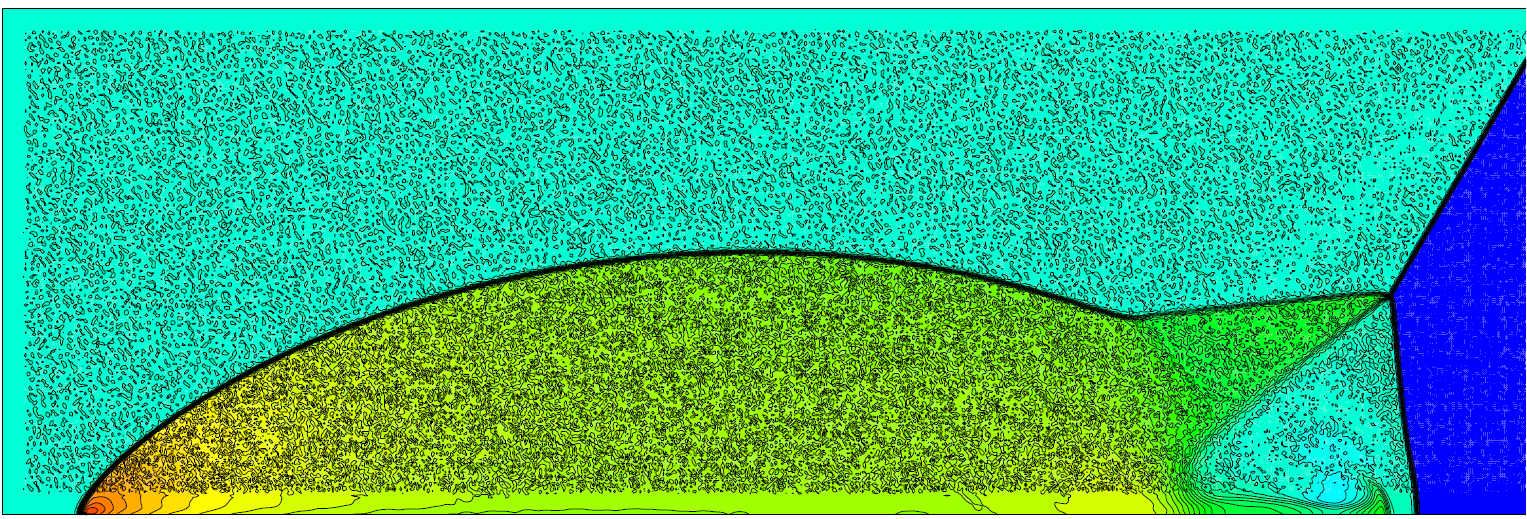}}
	\subfigure [ WENO-Like]{\includegraphics[width=0.75\textwidth]{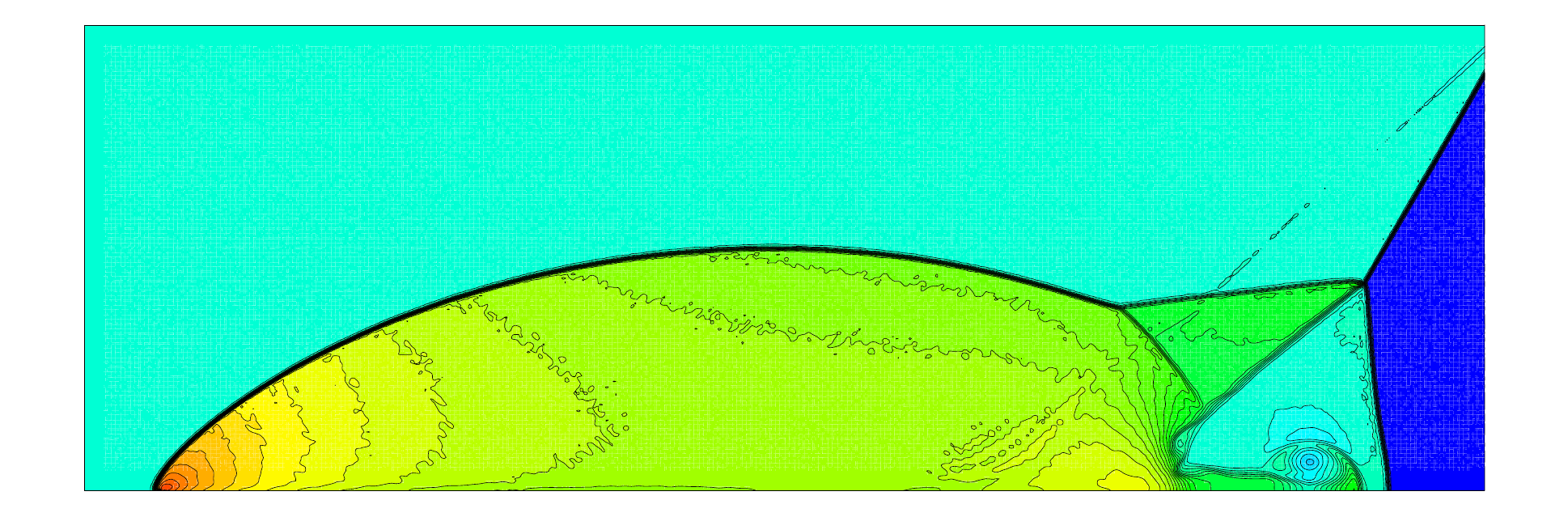}}
	\subfigure [ WENO-UFP]{\includegraphics[width=0.69\textwidth]{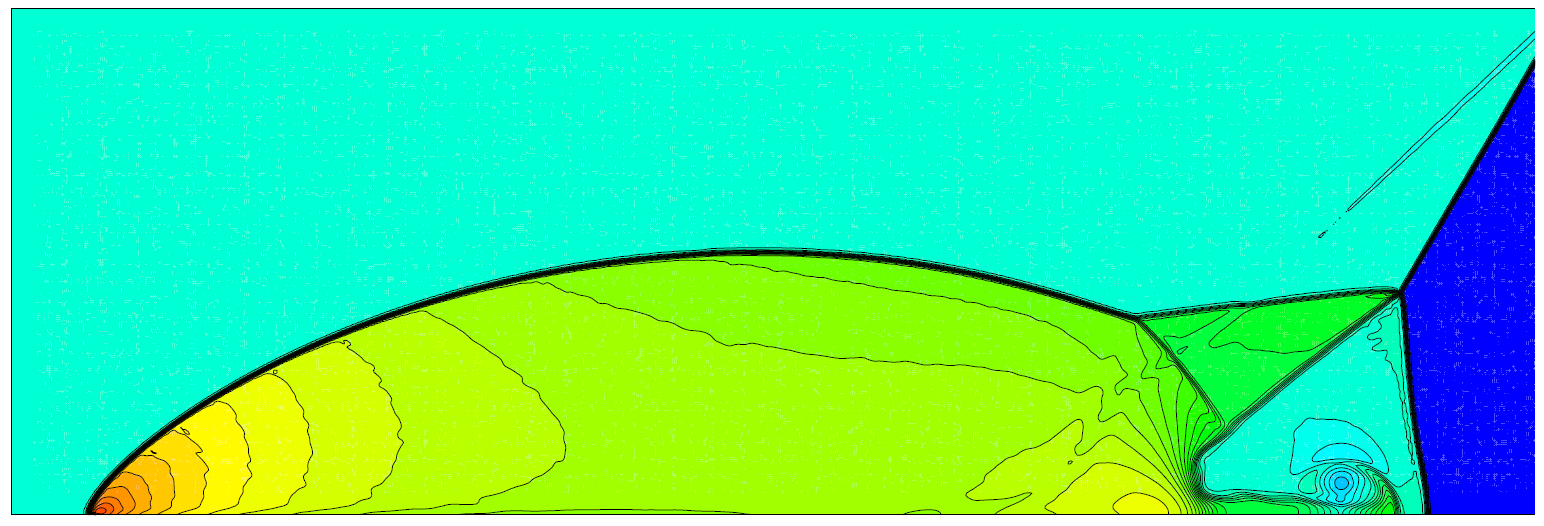}}
	\subfigure [ WENO-HUFP ] {\includegraphics[width=0.75\textwidth]{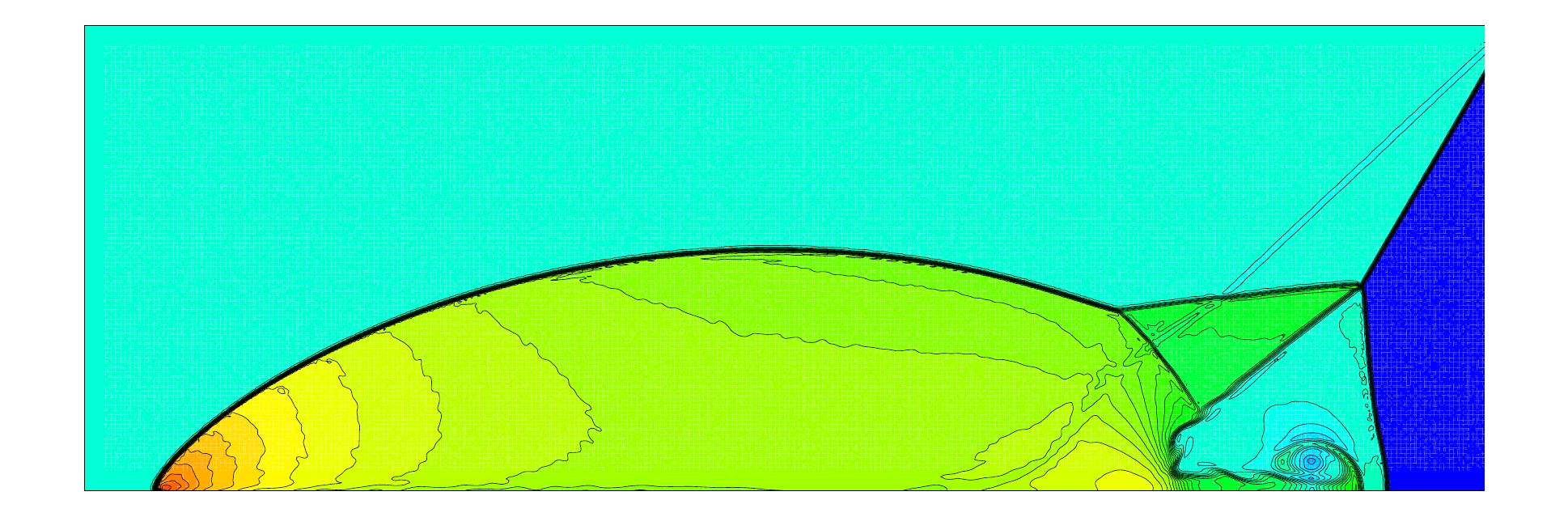}}
	\caption{  41 equally spaced density contours from 1.92 to 22.59 of 
		double Mach reflection problem on the $960\times 240$ grid.  }
	\label{DM-240}
\end{figure}

\begin{figure}\centering
	\subfigure [WENO]{\includegraphics[width=0.45\textwidth]{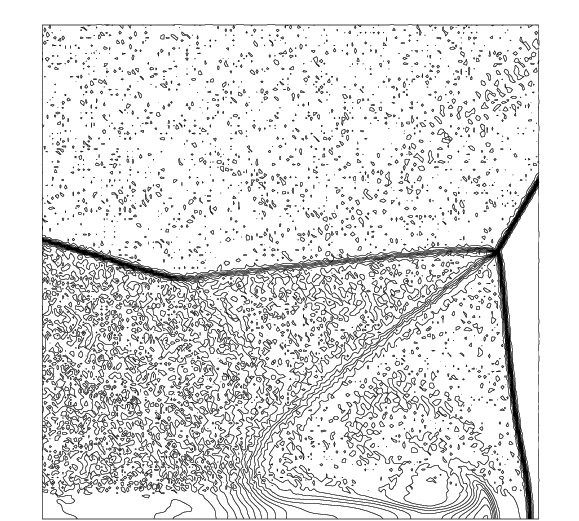}}
	\subfigure [ WENO-Like]{\includegraphics[width=0.45\textwidth]{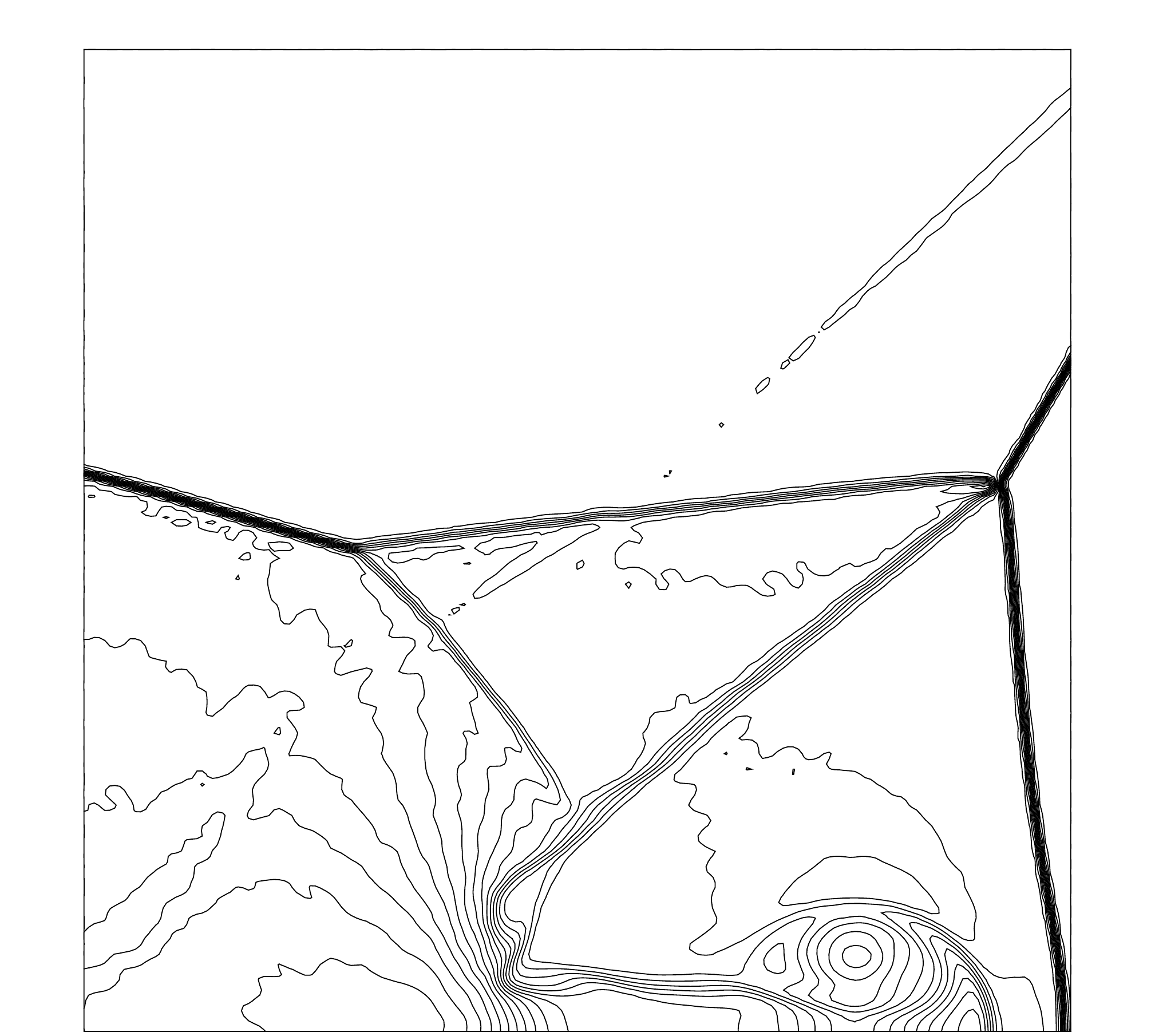}}
	\subfigure [ WENO-UFP]{\includegraphics[width=0.45\textwidth]{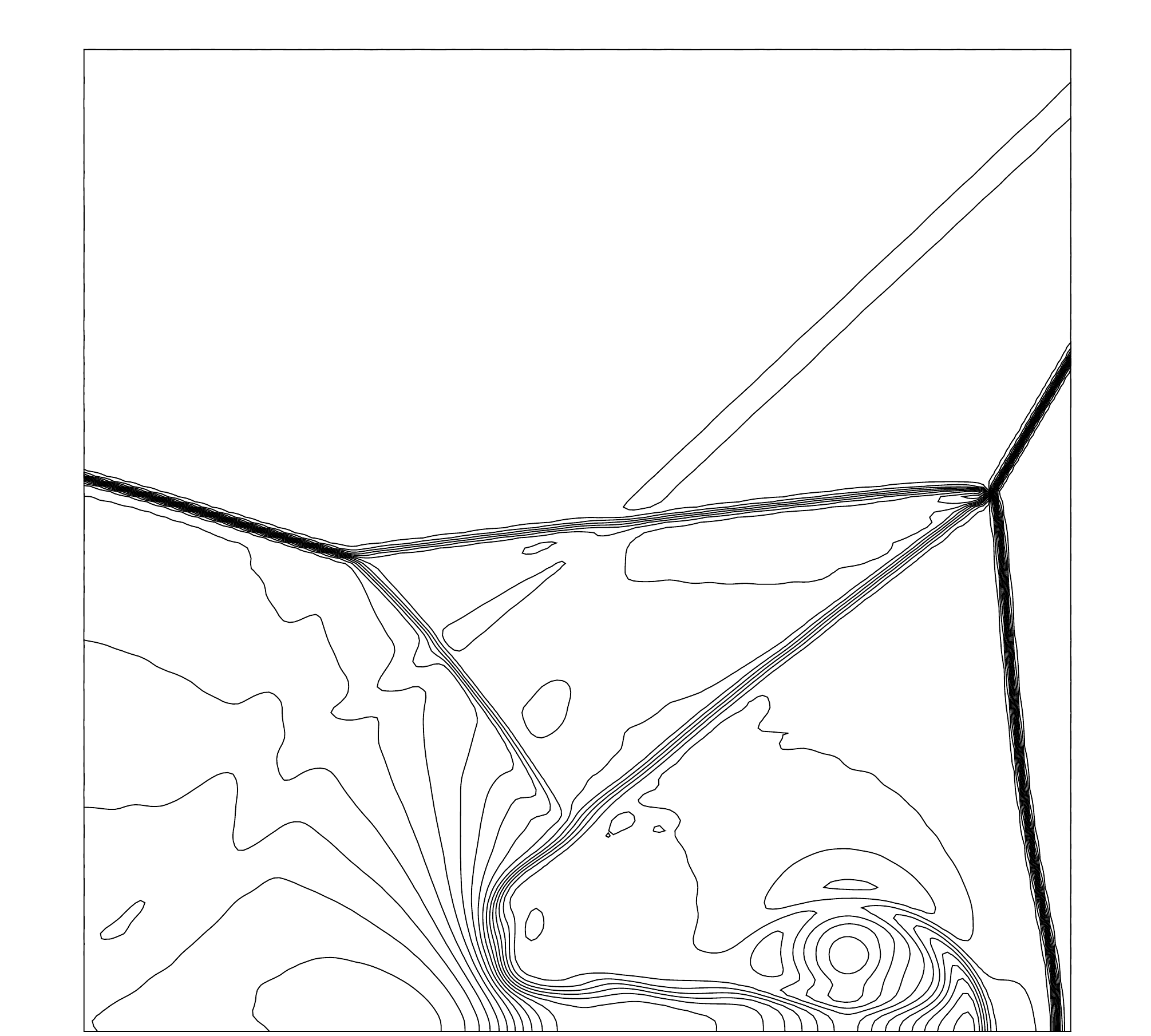}}
	\subfigure [ WENO-HUFP ] {\includegraphics[width=0.45\textwidth]{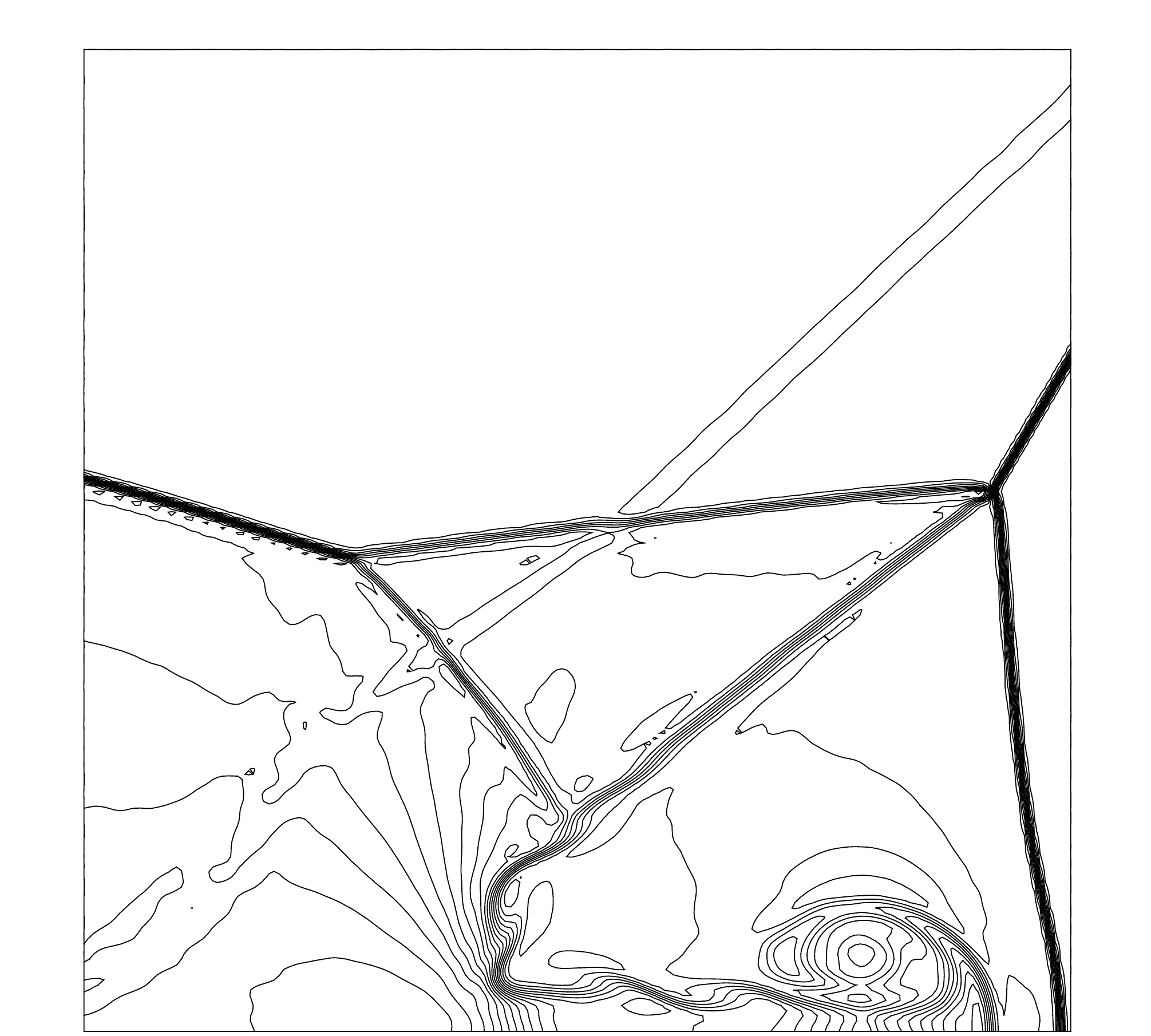}}
	\caption{  The enlarge part of double Mach reflection problem on $960\times 240$ grid.  }
	\label{DM-240-l}
\end{figure}

Since UPW5 is not able to resolve shock wave, their results are not shown here. 
From the density contours of the flow filed obtained on the $240\times 60$ grid, 
as shown in Fig.~\ref{DM-60}, 
it can be observed that WENO produces large errors in the region with grid perturbations. 
WENO-UFP and WENO-HUFP eliminate these errors and capture the shocks well 
and maintain the shock-capturing ability of the original WENO scheme on Cartesian grid. 
The density contours computed on the  $960\times 240$ grid 
and their enlarged part are shown in Fig.~\ref{DM-240} and Fig.~\ref{DM-240-l}, respectively. 
Here, WENO-Like denotes the method of Zhu et al.~\cite{zhu2017numerical}. 
Note that, WENO-HUFP is able to resolve more wave structures, as shown in Fig. \ref{DM-240-l}, 
than that of WENO-Like and WENO-UFP, which implies that WENO-HUFP is less dissipative. 
Also note that, while WENO-Like resolves less wave structre than WENO-HUFP, it produces considerable more fluctutions in the reflection wave region.
The computation time for different schemes are summrized in Table~\ref{double-mach-time}, 
it can be found that WENO-HUFP has the most computational efficiency 
and it costs less than two third of WENO-Like.

\begin{table}
	\scriptsize
	\centering
	\caption{Computational time for double Mach reflection problem on $960\times 240$ grid.}
	\begin{tabularx}{13.5cm}{@{\extracolsep{\fill}}llllr}
		\hline
		\quad & WENO &  WENO-Like &  WENO-UFP&  WENO-HUFP\\
		\hline
		CPU time (s) & $1923$	& $2951$& $2248$& $1834$ \\
		\hline
	\end{tabularx}
	\label{double-mach-time}
\end{table}  

\subsection{Supersonic flow past a cylinder}
\label{subsec1}
\begin{sidewaysfigure}\centering
	\subfigure [Grid]{\includegraphics[width=0.18\textwidth]{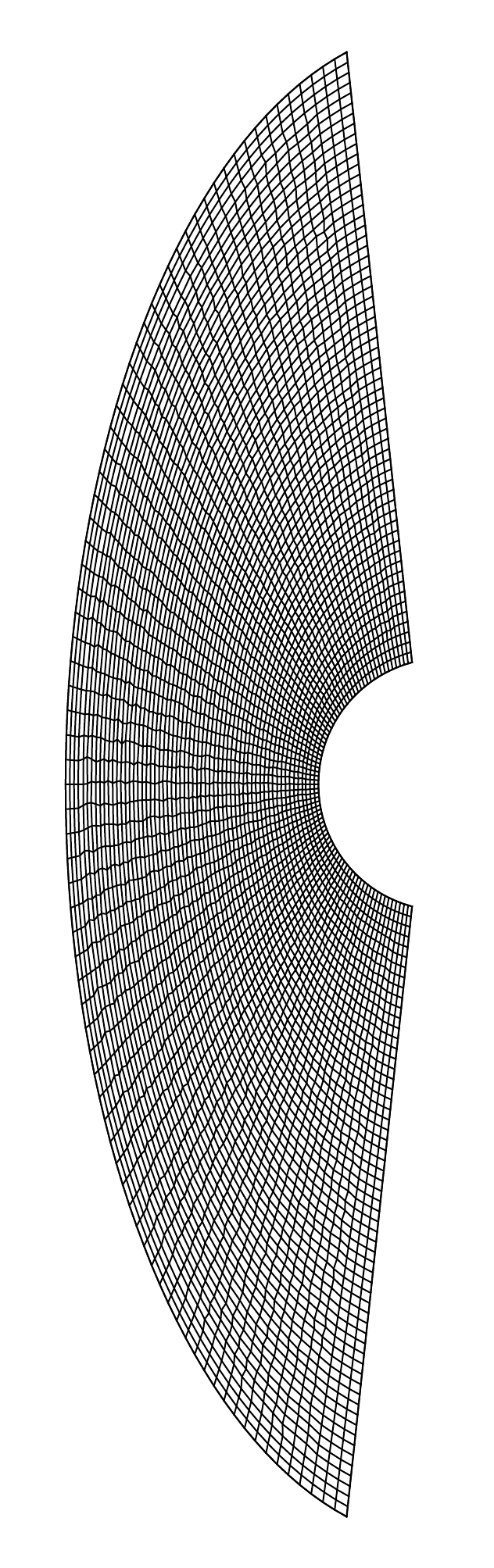}}
	\subfigure [ WENO]{\includegraphics[width=0.18\textwidth]{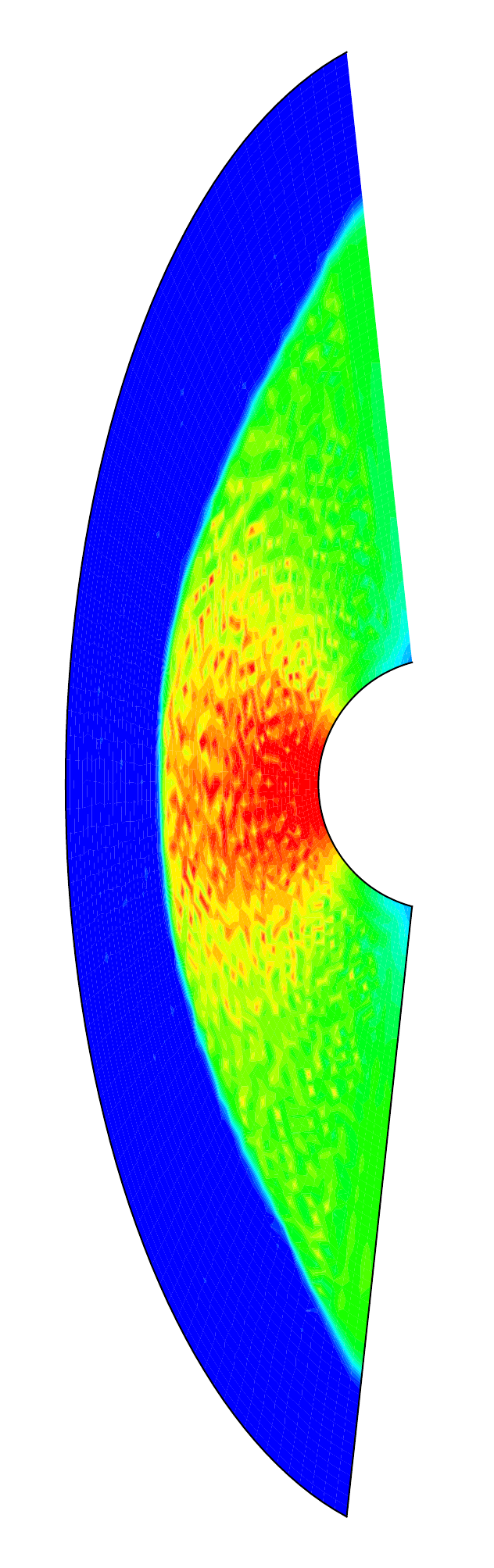}}
	\subfigure [ WENO-FP]{\includegraphics[width=0.18\textwidth]{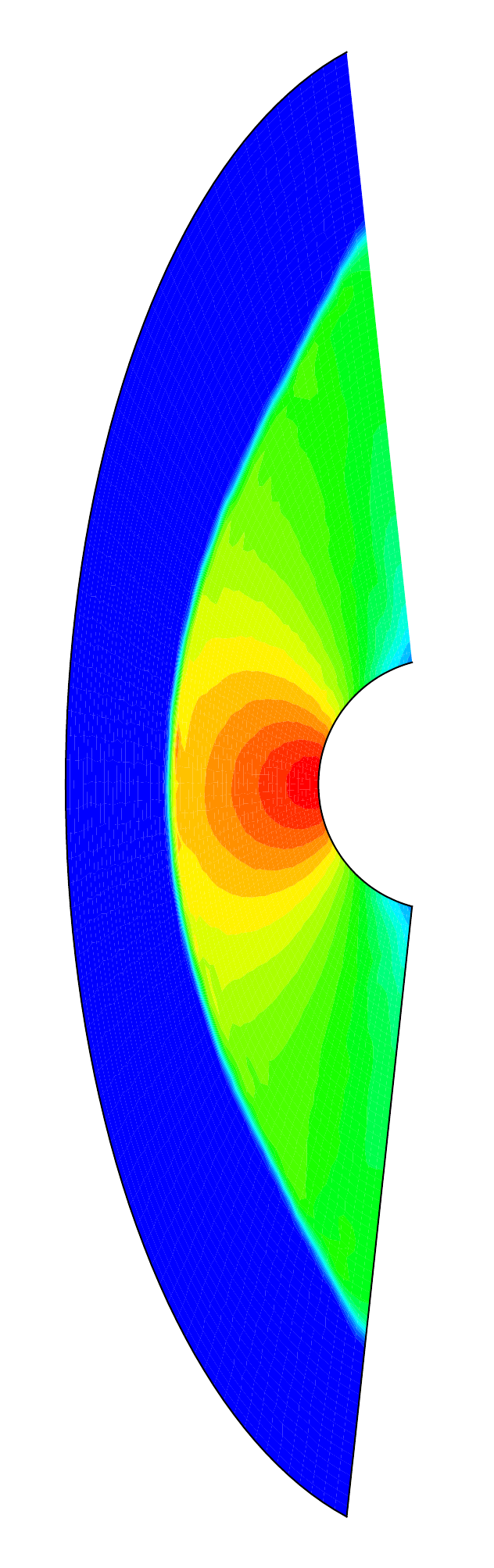}}
	\subfigure [ WENO-UFP ] {\includegraphics[width=0.18\textwidth]{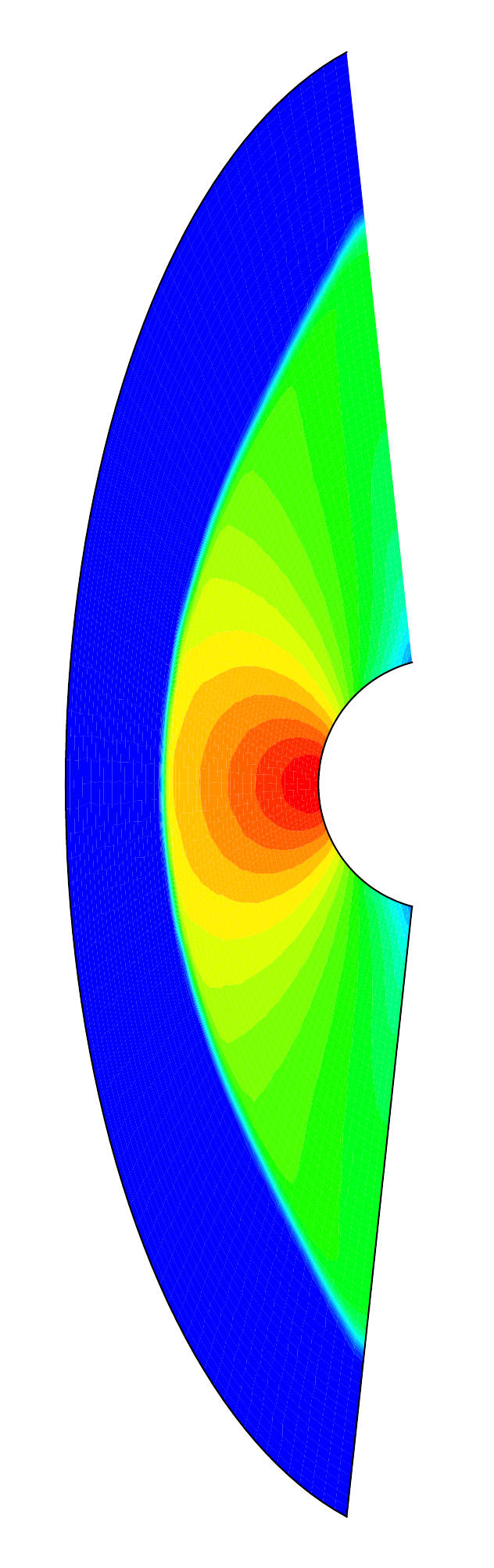}}
	\subfigure [ WENO-HUFP ] {\includegraphics[width=0.18\textwidth]{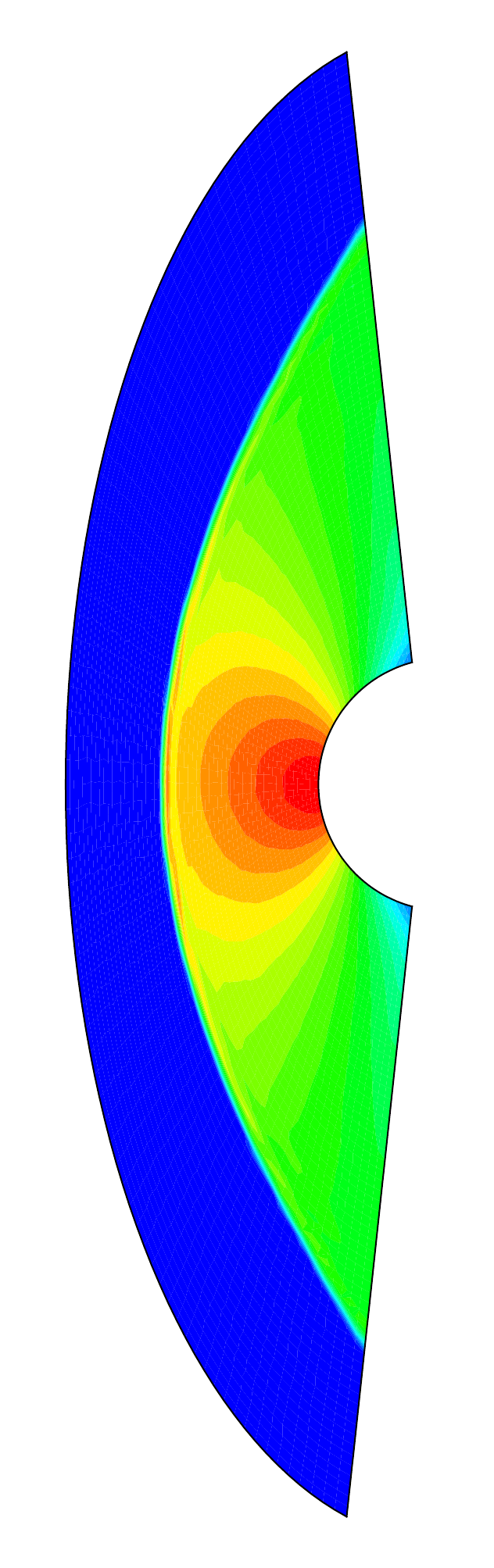}}
	\caption{ 21 equally spaced pressure contours from 0.86 to 3.88 of the supersonic flow past a cylinder.  }
	\label{cylinder}
\end{sidewaysfigure}

A supersonic flow past a cylinder~\cite{jiang1996efficient} is simulated on 
a grid randomized from a body-fitted grid. 
The Mach 2 supersonic flow with moves toward the cylinder from left. 
The reflective boundary condition is applied on the cylinder surface, 
the supersonic inflow condition at the left boundary 
and the supersonic outflow condition at the other boundaries. 
The gird with resolution of  $81\times 61$ is generated by the curvilinear coordinates
\begin{equation}
\begin{split}
&x=\left(R_x-\left( R_x-1\right)\eta'  \right) cos\left(\theta \left(2\xi'-1 \right)  \right)    \\
&y=\left(R_y-\left( R_y-1\right)\eta'  \right) sin\left(\theta \left(2\xi'-1 \right)  \right),    \\
\end{split} 
\label{eq-cy}
\end{equation}
where 
\begin{equation}
\begin{split}
&\xi'=\frac{\xi-1}{i_{max}-1}, \quad \xi=i+0.2\cdot \varphi_i   \\
&\eta'=\frac{\eta-1}{j_{max}-1}, \quad  \eta=j+0.2\cdot \sqrt{1-\varphi_i^2}. \\
\end{split} 
\label{eq-cy1}
\end{equation}
Here $\theta=5\pi / 12$, $R_x=3$ ,$R_y=6$ and $\varphi_i$ is a random number 
uniformly distributed between $\left[ 0,1\right] $. 
The inflow pressure and density are 
$\rho_{\infty}=1.0$ and $p_{\infty}=1/{\gamma}$, respectively. 
Similar to Ref.~\cite{WOODWARD1984115}, the time-step size is chosen 
as $\Delta t=0.005$ and the results are examined after $5000$ steps. 
The pressure contours and computational costs are given in Fig.~\ref{cylinder} 
and Table.~\ref{cylinder-time}, respectively. 
It is found that both WENO-FP and the present method eliminates the geometrically induced errors 
and maintain the shock capturing ability. 
Note that, WENO-HUFP is the least dissipative and the most computational efficient of all schemes. 

\begin{table}
	\scriptsize
	\centering
	\caption{Computational time for supersonic flow past a cylinder.}
	\begin{tabularx}{13.5cm}{@{\extracolsep{\fill}}llllr}
		\hline
		\quad & WENO &  WENO-FP &  WENO-UFP&  WENO-HUFP\\
		\hline
		CPU time (s) & $78$	& $109$& $104$& $62$ \\
		\hline
	\end{tabularx}
	\label{cylinder-time}
\end{table}

\section{Conclusions}
\label{sec1}
In this paper, 
we propose a free-stream preserving method for linear-upwind and WENO schemes on curvilinear grids. 
Following a Lax-Friedrichs flux splitting, 
the numerical fluxes of the upwind schemes are rewritten into a central term and 
a numerical dissipation term with the form of local difference 
using neighboring grid-point pairs. 
For the central term, the symmetric conservative metric method is applied straightforwardly 
to eliminate the geometrically induced error. 
For the numerical dissipation term, 
each neighboring grid-point pairs are modified to share a common Jacobian and metrics value 
which are evaluated by high order schemes.  
Then, a simple free-stream preserving hybrid method switching between linear-upwind and WENO schemes 
is proposed to further improve computational efficiency and reduce numerical dissipation. 
A number of numerical examples demonstrate that the proposed method not only achieves 
good free-stream and vortex preserving properties 
but also maintains the shock-capturing ability of original WENO scheme. 
In addition, the hybrid method achieves higher resolution solution 
than those of WENO-UFP and WENO-Like schemes 
with considerable lower computing costs.



\section*{Acknowledgements}
\addcontentsline{toc}{section}{Acknowledgement}
The first author is partially supported by Xidian University (China).
The second author acknowledges 
National Natural Science Foundation of China(NSFC) (Grant No:11628206).


\bibliographystyle{elsarticle-num}

\bibliography{GCL}

\end{document}